\documentclass[12pt]{article}
\usepackage{epsfig,amssymb,amsmath,psfrag}

\def \bl  {\begin{align*}}
\def \el  {\end{align*}}

\def \be  {\begin{equation}}
\def \ee  {\end{equation}}
\def \ba  {\begin{eqnarray}}
\def \ea  {\end{eqnarray}}
\def \baa {\begin{eqnarray*}}
\def \eaa {\end{eqnarray*}}
\def \bb  {\begin {thebibliography} }
\def \eb  {\end{thebibliography}}
\def \lab #1 {\label{#1}}

\def \qqquad {\qquad\quad}

\def \matrix #1 {\left(\begin{array}{cc} #1 \end{array}\right)}

\renewcommand{\a}{\alpha}
\newcommand{\da}{{\dot{\alpha}}}
\newcommand{\db}{{\dot{\beta}}}
\renewcommand{\b}{\beta}

\newcommand\lr[1]{{\left({#1}\right)}}
\newcommand \widebar [1] {\overline{#1}}

\newcommand \vev [1] {\langle{#1}\rangle}

\def\XXint#1#2#3{{\setbox0=\hbox{$#1{#2#3}{\int}$}
     \vcenter{\hbox{$#2#3$}}\kern-.5\wd0}}

\def\l<{\langle}\def\r>{\rangle}
\def\bMHV{\overline{\rm MHV}}

\begin{document}

\thispagestyle{empty}
\null\vskip-12pt \hfill  LAPTH-1267/08 \\
\vskip2.2truecm
\begin{center}
\vskip 0.2truecm {\Large\bf
{\Large All tree-level amplitudes in $\mathcal{N}=4$ SYM}
}\\
\vskip 1truecm
{\bf J.~M. Drummond$^{*}$ and J.~M. Henn$^{*}$ \\
}

\vskip 0.4truecm
$^{*}$ {\it
LAPTH\footnote{Laboratoire d'Annecy-le-Vieux de Physique Th\'{e}orique, UMR 5108}, Universit\'{e} de Savoie, CNRS\\
B.P. 110,  F-74941 Annecy-le-Vieux Cedex, France\\
\vskip .2truecm                        }
\end{center}

\vskip 1truecm 
\centerline{\bf Abstract} 
We give an explicit formula for all tree amplitudes in $\mathcal{N}=4$ SYM, derived by solving the recently presented supersymmetric tree-level recursion relations.
The result is given
in a compact, manifestly supersymmetric form and we show how to extract from it all possible component amplitudes for an arbitrary number of external particles and any arrangement of external particles and helicities. We focus particularly on extracting  gluon amplitudes which are valid for any gauge theory.
The formula for all tree-level amplitudes is given in terms of nested sums of dual superconformal invariants and it therefore manifestly respects both conventional and dual superconformal symmetry.

\medskip

 \noindent

\newpage
\setcounter{page}{1}\setcounter{footnote}{0}


\section{Introduction}

Gluon scattering amplitudes are known to have many remarkable
properties. In a recent paper \cite{dhks5}, it was discovered that
in $\mathcal{N}=4$ SYM, scattering amplitudes exhibit a new, dual
superconformal symmetry. This new symmetry appears in addition to all
previously known symmetries of the amplitudes.
It was also shown that this dual superconformal symmetry can
be understood through the AdS/CFT correspondence, where it
appears as a symmetry of the $AdS_{5}\times S^{5}$ string
sigma model \cite{Berkovits:2008ic,Beisert:2008iq}. In this paper we
will construct a solution for all tree-level amplitudes in
$\mathcal{N}=4$ SYM and show explicitly how it respects dual
superconformal symmetry.

The first hint at an unexpected simplicity in gluon scattering
amplitudes was the formula for the MHV amplitudes conjectured by Parke
and Taylor
\cite{Parke:1986gb} (and later proved by Berends and Giele
\cite{Berends:1987me}). For amplitudes having generic helicity
configurations, Witten argued that they have remarkable properties
in twistor space \cite{Witten:2003nn}. This conjecture was
verified for NMHV amplitudes \cite{Britto:2004tx,Bern:2004bt},
however the explicit formulae \cite{Kosower:2004yz} for these amplitudes are
rather complicated. Since tree
level gluon amplitudes in $\mathcal{N}=4$ SYM are equal to gluon
amplitudes in any gauge theory, including QCD, it is no
restriction to consider amplitudes in $\mathcal{N}=4$ SYM instead.
Keeping this in mind and having observed that $\mathcal{N}=4$ SYM
amplitudes have an additional symmetry, dual superconformal
symmetry, it seems natural to write the amplitudes in a manifestly
supersymmetric way. The appropriate on-shell $\mathcal{N}=4$
superspace was introduced by Nair \cite{Nair:1988bq}, who
used it to write down the MHV super-amplitudes. This superspace was employed in \cite{Witten:2003nn} to describe amplitudes in super-twistor space and in \cite{Georgiou:2004by} to express NMHV amplitudes using a supersymmetric version of the CSW rules \cite{Cachazo:2004kj}. Employing this
superspace will allow us to make the additional symmetry
properties of the amplitudes manifest and hopefully lead
to simpler expressions than the previously available ones.
Indeed, it was conjectured \cite{dhks5} and later proved \cite{dhks6}
that NMHV tree level amplitudes written in this
superspace have a remarkably simple form, they are just given by a
sum over certain dual superconformal invariants. It seems natural
to expect that one can go beyond NMHV amplitudes and that generic
N${}^{p}$MHV amplitudes will have a relatively simple form when
written in superspace. Since these super-amplitudes are not yet
known we compute them in this paper.

The state-of-the-art method for computing tree-level scattering
amplitudes in gauge theory are the BCF/BCFW on-shell recursion
relations \cite{Britto:2004ap,Britto:2005fq}. Recently, these
recursion relations have been written for $\mathcal{N}=4$ SYM in
on-shell superspace \cite{Bianchi:2008pu,arkani,Brandhuber:2008pf,ArkaniHamed:2008gz,Elvang:2008na}.
We will use the form presented in \cite{arkani,Brandhuber:2008pf,ArkaniHamed:2008gz}.
This is precisely the
tool we need to study tree-level super-amplitudes for arbitrary
helicity configurations.
The supersymmetric recursion relations have been used very recently
to verify that tree-level scattering amplitudes in $\mathcal{N}=4$ SYM are covariant
under dual superconformal transformations \cite{Brandhuber:2008pf}.

In this paper, we use the supersymmetric recursion relations to
compute tree-level amplitudes in $\mathcal{N}=4$ SYM. As we will
see, writing the recursion relations in superspace makes it
significantly simpler to solve them.
We use the explicit solutions for NMHV, NNMHV, and NNNMHV amplitudes as examples to study the general pattern and then
we present a solution for all amplitudes in terms of nested sums.
Our result on NMHV amplitudes confirms the result of \cite{dhks6}, while our
results for generic non-MHV amplitudes are new.

We then study the symmetries of our solution and show how the
conventional superconformal symmetry of $\mathcal{N}=4$ SYM is
realised on the amplitudes. We also study the dual superconformal
symmetry that the tree-level super-amplitudes should exhibit
\cite{dhks5}. This symmetry is a generalisation of dual conformal
symmetry, which first appeared as a property of loop integrals in the
perturbative expansion of MHV amplitudes
\cite{Drummond:2006rz,Bern:2006ew,Bern:2007ct} and then, in the
context of the AdS/CFT correspondence, as the isometry of a T-dual
AdS${}_5$ in \cite{Alday:2007hr,Alday:2007he} and finally as an
anomalous Ward identity for MHV amplitudes
\cite{Drummond:2007cf,Drummond:2007au}. This last manifestation of
dual conformal symmetry is based on a conjectured duality between MHV
amplitudes and Wilson loops
\cite{Alday:2007hr,Drummond:2007aua,Brandhuber:2007yx} which has been
confirmed in perturbation theory up to two loops
\cite{Drummond:2007cf,Drummond:2007au,Drummond:2007bm,Bern:2008ap,Drummond:2008aq}.
A review of these developments is given in \cite{Alday:2008yw}.

The paper is organised as follows. In section
\ref{sect-technicalities} we introduce the necessary superspace
definitions and briefly review the extension of the BCF recursion relations
to superspace. In section \ref{sect-NMHV}, we show how to solve the supersymmetric
recursion relations in the NMHV case, and in section
\ref{sect-NNMHV} in the NNMHV case. Based on the previous sections, we
give in section \ref{sect-all} the solution to the supersymmetric relations for the generic non-MHV case.
In section \ref{sect-app-symmetry} we discuss both the conventional and dual
superconformal symmetry of our solutions.
Section \ref{sect-app-components} serves to explain how to
extract gluon scattering amplitudes from our super-amplitudes.
Section \ref{sect-conclusions} contains our conclusions.
There are two appendices. In appendix \ref{app-collinear} we discuss the behaviour of
our results under the collinear limit.
In appendix \ref{app-generators} we give the generators of the ordinary as well as the dual
superconformal algebra.

\section{Amplitudes and supersymmetric recursion relations}\label{sect-technicalities}

In this paper, we will be discussing colour-ordered scattering amplitudes.
The tree-level MHV gluon amplitudes mentioned in the introduction
are given by \cite{Parke:1986gb,Berends:1987me} \footnote{In this
paper we omit the standard factor of $i (2 \pi)^4$ in the
normalisation of the amplitudes.} \be \label{intro-MHV-gluons}
A(1^{-},2^{+},\ldots,j^{-},\ldots,n^{+}) =   \delta^{(4)}(p)\,
\frac{\langle 1 j\rangle^4}{\vev{1\, 2}\vev{2\, 3}\ldots\vev{n\,
1}}\,, \ee where $p=\sum_{i=1}^n\ \lambda_{i}^{\alpha}\,
    \tilde\lambda_{i}^{\dot\alpha}$ is the total momentum and $\vev{ij}=\lambda_{i}^{\alpha} \lambda_{j\,\alpha}$.
In order to shed more light on gluon scattering amplitudes of
arbitrary helicity configurations and make their symmetries
manifest, it is useful to consider scattering amplitudes in
$\mathcal{N}=4$ SYM, which has many exceptional properties.
Using Grassmann variables $\eta^{A}$ we can write down a super-wavefunction
\ba \label{super-wave}
  \Phi(p,\eta) &=& G^{+}(p) + \eta^A \Gamma_A(p) + \frac{1}{2}\eta^A \eta^B S_{AB}(p)
  + \frac{1}{3!}\eta^A\eta^B\eta^C \epsilon_{ABCD} \bar\Gamma^{D}(p) \nonumber \\
  &&\  + \frac{1}{4!}\eta^A\eta^B\eta^C \eta^D \epsilon_{ABCD} G^{-}(p)\,,
\ea
which incorporates as its components all on-shell states of $\mathcal{N}=4$ SYM.
Since the $\mathcal{N}=4$ supermultiplet is PCT self-conjugate, we could equally well have chosen
an anti-chiral representation (see \cite{dhks5,dhks6} for more explanations).
Then we can define super-amplitudes as
\be\label{super-amplitude}
{\cal A}_n\big(\lambda,\tilde{\lambda},\eta\big) = \mathcal{A}\left( \Phi_1 \ldots \Phi_n \right)\,.
\ee
In this paper we will be discussing exclusively tree-level amplitudes.
The $\mathcal{N}=4$  supersymmetric version of the MHV tree-level amplitude
(\ref{intro-MHV-gluons}) then reads \cite{Nair:1988bq} \be
\label{intro-MHV-n}
  {\cal A}^{\rm MHV}_{n}(\lambda,\tilde\lambda,\eta) = \frac{ \delta^{(4)}(p)\,
    \delta^{(8)} (q)}{\vev{1\, 2}\vev{2\, 3}\ldots\vev{n\, 1}}\,,
\ee where $q=\sum_{i=1}^n\ \lambda_{i}^{\alpha}\, \eta^A_i$. The
appearance of $ \delta^{(8)} (q)$ is dictated by ${\mathcal N}=4$
supersymmetry, and can be thought of as imposing super-momentum
conservation, just as $\delta^{(4)}(p)$ ensures momentum
conservation.

The full tree-level super-amplitude (\ref{super-amplitude}) contains not just MHV but all possible N${}^p$MHV super-amplitudes and has the factors $\delta^{(4)}(p)$ and $\delta^{(8)}(q)$ for the same reason. It is convenient to factor out the MHV tree-level super-amplitude (\ref{intro-MHV-n}) and write the remaining factor as $\mathcal{P}_n$,
\be
\label{def-curlyP}
\mathcal{A}_n = \mathcal{A}_n^{\rm MHV} \mathcal{P}_n.
\ee
The factor $\mathcal{P}_n$ has an expansion in the Grassmann parameters $\eta$,
\be
\mathcal{P}_n = \mathcal{P}_n^{\rm MHV} + \mathcal{P}_n^{\rm NMHV} + \ldots \mathcal{P}_n^{\bMHV}.
\ee
Of course $\mathcal{P}_n^{\rm MHV} =1$ while $\mathcal{P}_n^{\rm NMHV}$ has Grassmann degree 4 and the remaining terms increase in Grassmann degree in units of 4 up to $\mathcal{P}_n^{\bMHV}$ which is of degree $4n-16$.

The super-amplitude ${\cal A}^{\rm MHV}_{n}$
contains the pure gluon amplitude (\ref{intro-MHV-gluons}) as a
component in the expansion in the Grassmann parameters $\eta_{i}$,
\be \label{intro-expansion} {\cal A}^{\rm MHV}_{n} =
\left(\eta_1\right)^4 \left(\eta_j\right)^4
A(1^{-},2^{+},\ldots,j^{-},\ldots,n^{+}) + \ldots\,, \ee where
$(\eta)^4 = (1/4!) \epsilon_{ABCD} \eta^A \eta^B \eta^C \eta^D$.
The full super-amplitude $\mathcal{A}_n$ contains all gluon amplitudes (with arbitrary total helicity) as well as all amplitudes with fermions and scalars in $\mathcal{N}=4$ SYM.
The
superspace formulation of the amplitudes has the advantage that
supersymmetric Ward identities are automatically satisfied.
Another advantage is that, as was conjectured in \cite{dhks5} and
proved in \cite{dhks6}, NMHV amplitudes have a particularly simple
form when written in superspace, namely
\be\label{intro-NMHV}
 {\cal A}^{\rm {NMHV}}_{n} = \mathcal{A}_n^{\rm MHV} \mathcal{P}_n^{\rm NMHV} =\frac{ \delta^{(4)}(p) \, \delta^{(8)}(q)}{\vev{1\, 2}\vev{2\, 3}\ldots\vev{n\, 1}} \, \sum_{1<s<t<n} R_{n;st} \,,
\ee
where $R_{n;st}$ are dual superconformal invariants
whose precise form is given in \cite{dhks5} and will be given again shortly.

\begin{figure}
\psfrag{dots}[cc][cc]{$\ldots$}
\psfrag{one}[cc][cc]{$\hat{1}$}
\psfrag{two}[cc][cc]{$2$}
\psfrag{three}[cc][cc]{$3$}
\psfrag{en}[cc][cc]{$\bar{n}$}
\psfrag{bigen}[cc][cc]{$N$}
\psfrag{pe}[cc][cc]{$\hat{P}_{i}$}
\psfrag{i}[cc][cc]{\hspace{0.5cm}$i-1$}
\psfrag{iplus}[cc][cc]{$i$}
\psfrag{sumi}[cc][cc]{$\sum$}
\psfrag{AL}[cc][cc]{$\mathcal{A}_{\rm L}$}
\psfrag{AR}[cc][cc]{$\mathcal{A}_{\rm R}$}
\psfrag{x1t1}[cc][cc]{$\hat{x}_{1}\,,\hat{\theta}_{1}$}
\psfrag{x2t2}[cc][cc]{${x}_{2}\,,{\theta}_{2}$}
\psfrag{xiti}[cc][cc]{${x}_{i}\,,{\theta}_{i}$}
\psfrag{ximtim}[cc][cc]{${x}_{i-1}\,,{\theta}_{i-1}$}
\psfrag{xnmtnm}[cc][cc]{${x}_{n}\,,{\theta}_{n}$}
\psfrag{BCF}[cc][cc]{r.h.s. of on-shell recursion relation}
\psfrag{dualBCF}[cc][cc]{dual variables}
 \centerline{{\epsfysize4cm
\epsfbox{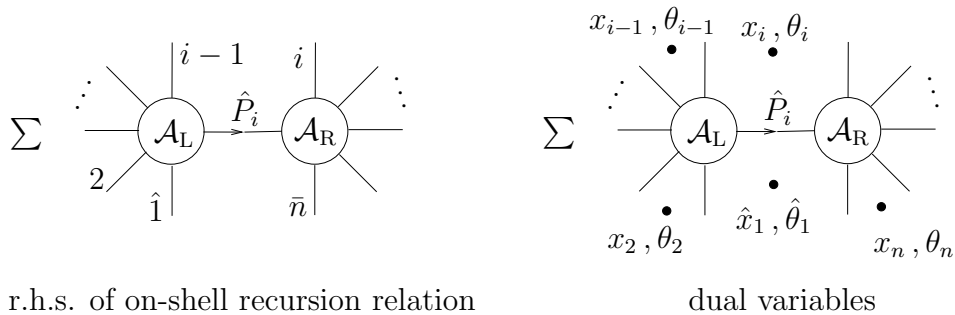}}}  \caption[]{\small
  Illustration of the r.h.s of the on-shell recursion relations (\ref{BCF}),(\ref{BCF-super}). The picture on the right illustrates the transition to dual variables.}
  \label{fig-bcf}
\end{figure}

Let us now quickly introduce the necessary information on the BCF
on-shell recursion relations. They express $n$-point scattering
amplitudes in terms of a sum over a product of scattering
amplitudes of fewer points \cite{Britto:2004ap,Britto:2005fq}.
Schematically, they read 
\be \label{BCF} \mathcal{A} =
\sum_{P_{i}} \sum_{h} \mathcal{A}_{L}^{h}(z_{P_{i}})
\frac{1}{P_{i}^2} \mathcal{A}_{R}^{-h}(z_{P_{i}})\,. 
\ee
In
(\ref{BCF}), $z_{P}$ indicates that in the amplitudes on the r.h.s
certain momenta were shifted. The shift can be chosen in many ways. For our purposes it is convenient to shift two adjacent legs according to 
\be\label{shift1}
\hat{\tilde{\lambda}}_{n} = \tilde{\lambda}_{n} + z_{P_{i}}
\tilde{\lambda}_{1}\,, \qquad \hat{\lambda}_{1} = \lambda_{1} -
z_{P_{i}} \lambda_{n}\,. 
\ee 
Hatted quantities denote the shifted
variables. This shift, called an $|n 1\rangle$ shift, is depicted
in Fig. \ref{fig-bcf}. Note that the amplitudes
$\mathcal{A}^{h}_{L}(z_{P_{i}}),\mathcal{A}^{-h}_{R}(z_{P_{i}})$
are on-shell. Indeed, the shift parameter $z_{P}$ must be chosen
such that this is the case, which amounts to saying that the
shifted intermediate momentum $\hat{P}_{i} = - (\hat{\lambda}_{1}
\tilde{\lambda}_{1} + \sum_{j=2}^{i-1} \lambda_j
\tilde{\lambda}_{j} )$ is on-shell, i.e. 
\be 
(\hat{P}_{i})^2 =
\left( - \sum_{j=1}^{i-1} \lambda_j \tilde{\lambda}_{j} + z_{P_{i}} \lambda_{n} \tilde{\lambda}_{1} \right)^2 =  0\,.\\
\ee 
Note also that the propagator $1/P_{i}^2$ in (\ref{BCF}) is evaluated for
unshifted kinematics.

We will use the supersymmetric version of the BCF recursion relations of \cite{arkani,Brandhuber:2008pf,ArkaniHamed:2008gz}.
This amounts to replacing the sum over intermediate states by a superspace
integral, and the on-shell amplitudes by super-amplitudes, i.e.
\be \label{BCF-super} \mathcal{A} = \sum_{P_{i}} \int
d^{4}\eta_{P_{i}} \mathcal{A}_{L}(z_{P_{i}}) \frac{1}{P_{i}^2}
\mathcal{A}_{R}(z_{P})\,. \ee The validity of the supersymmetric
equations can be justified by relating the $z \to \infty$
behaviour of the shifted super-amplitudes $\mathcal{A}(z)$ to the
known behaviour of component amplitudes \cite{Britto:2005fq}
using supersymmetry \cite{arkani,Brandhuber:2008pf,ArkaniHamed:2008gz}.

For the supersymmetric equations, supersymmetry requires that in
addition to (\ref{shift1}) we also have \be \label{shift2}
\hat{\eta}_{n} = \eta_{n} + z_{P_{i}} \eta_{1} \,. \ee
In the following sections it will be very useful to use the dual variables \cite{Drummond:2006rz}
\be \label{dualx}
\lambda_{i}\tilde\lambda_{i}   = x_{i} - x_{i+1}\,.
\ee
As was already mentioned, these have a natural generalisation to dual superspace \cite{dhks5}, i.e.
\be \label{dualth}
\lambda_{i} \eta_{i}  = \theta_{i} - \theta_{i+1}\,.
\ee
Following \cite{Brandhuber:2008pf}, in the supersymmetric recursion relations only the following dual
variables get shifted,
\be
\hat{x}_{1} = x_{1} - z_{P_{i}} \lambda_{n} \tilde{\lambda}_{1}\,,\qquad \hat{\theta}_{1} = \theta_{1} - z_{P_{i}} \lambda_{n} \eta_{1}\,.
\ee
See Fig. \ref{fig-bcf}.
The fact that all other dual variables remain inert under the shift will prove useful when solving the supersymmetric recursion relations.

\section{NMHV tree amplitudes}\label{sect-NMHV}

Here we show that it is straightforward to obtain all NMHV tree amplitudes from the supersymmetric recursion relation (\ref{BCF-super}) and knowing the MHV super-amplitudes.

\begin{figure}
\psfrag{dots}[cc][cc]{$\ldots$}
\psfrag{one}[cc][cc]{$\hat{1}$}
\psfrag{two}[cc][cc]{$2$}
\psfrag{three}[cc][cc]{$3$}
\psfrag{en}[cc][cc]{$\bar{n}$}
\psfrag{bigen}[cc][cc]{$N$}
\psfrag{pe}[cc][cc]{$\hat{P}$}
\psfrag{pei}[cc][cc]{$\hat{P}_{i}$}
\psfrag{i}[cc][cc]{\hspace{0.5cm}$i-1$}
\psfrag{iplus}[cc][cc]{$i$}
\psfrag{A}[cc][cc]{$A$}
\psfrag{B}[cc][cc]{$B$}
\psfrag{sumi}[cc][cc]{$\sum_{i=4}^{n-1}$}
 \centerline{{\epsfysize3cm
\epsfbox{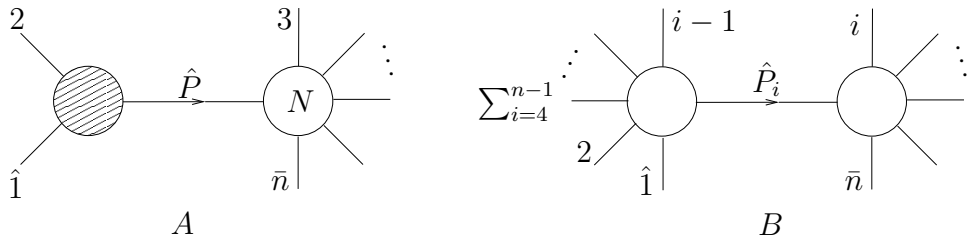}}}  \caption[]{\small
  The two contributions to the supersymmetric recursion relation for
  NMHV amplitudes. We call term $B$ inhomogeneous and $A$
  homogeneous. $B$ can be easily computed since it is built from MHV
  amplitudes only. $\hat{1}$ means that $\lambda_{1}$ is shifted, and
  $\bar{n}$ means that $\tilde{\lambda}_{n}$ is shifted.}
  \label{fig1}
\end{figure}

Apart from the $n$-point MHV super-amplitude (\ref{intro-MHV-n})
we need the $3$-point $\widebar{\rm MHV}$ amplitude, which can be
readily obtained from (\ref{intro-MHV-n}) for $n=3$ by a Grassmann
Fourier transform and complex conjugation,
\be
\label{bar-MHV3}
  {\cal A}^{\rm \widebar{MHV}}_{3}(\lambda,\tilde\lambda,\eta) = \delta^{(4)}(p)
  \frac{\delta^{(4)}\lr{\eta_1[23]+\eta_2[31]+\eta_3[12]}}{[12][23][31]}\,.
\ee 
The form of the three-point $\bMHV$ amplitude has appeared already in \cite{arkani,Brandhuber:2008pf,ArkaniHamed:2008gz,dhks6}.
NMHV super-amplitudes have Grassmann degree $12$. Looking at
(\ref{BCF-super}) we see that there is a Grassmann integration,
which means that the Grassmann degree of the amplitudes on the
r.h.s. of (\ref{BCF-super}) must add up to $16$. This is only
possible in two ways, $4+12$ and $8+8$, which corresponds to
taking $\widebar{\rm MHV}_{3} + {\rm NMHV}$ and ${\rm MHV} + {\rm
MHV}$ amplitudes for $\mathcal{A}_{L},\mathcal{A}_{R}$,
respectively. It is convenient to choose a shift of two
neighbouring points, e.g. a $\lbrack n 1 \rangle$ shift. Then the
supersymmetric recursion relation for $\mathcal{A}_{n}^{\rm NMHV}$
reads
\begin{align}
\label{super-BCF-NMHV}
\mathcal{A}_{n}^{\rm NMHV} &= \int
\frac{d^{4}P}{P^2} \int d^{4} \eta_{\hat{P}} \,
\mathcal{A}_{3}^{\rm \widebar{MHV}}(z_{P}) \mathcal{A}_{n-1}^{\rm
NMHV}(z_{P}) + \sum_{i=4}^{n-1}  \int \frac{d^{4}P_{i}}{P_{i}^2}
\int d^{4}\eta_{\hat{P}_{i}} \mathcal{A}_{i}^{\rm MHV}(z_{P_{i}}) 
\mathcal{A}_{n-i+2}^{\rm MHV}(z_{P_{i}})\notag\\
&\equiv A + B\, .
\end{align}
The two terms in
(\ref{super-BCF-NMHV}) are depicted in Fig. \ref{fig1}.

Note that the shifted lines must be on opposite sides of the
exchanged line. Note also that the leg $n$ with the
anti-holomorphic shift cannot connect to the $\widebar{\rm
MHV}_{3}$ amplitude since this would not be allowed by the
kinematics. Similarly, an ${\rm MHV}_{i}$ amplitude containing the
leg $1$ with the holomorphic shift must have at least four
legs, which explains the range of $i$ in
(\ref{super-BCF-NMHV}).

\subsection{Inhomogeneous term}
\label{subsect-inhom}
The inhomogeneous term in the recursion relation
(\ref{super-BCF-NMHV}) for NMHV amplitudes (corresponding to Fig.
\ref{fig1}\,B) can be readily calculated since it is built
entirely from the known MHV amplitudes, see (\ref{intro-MHV-n}).

By writing, for example, the Grassmann delta function coming from
$\mathcal{A}_{i}^{\rm MHV}(z_{P})$ in the following way,
\be
\delta^{(8)}\left(\hat{\lambda}_{1} \eta_{1} + \sum_{j=2}^{i-1}
\lambda_{j} \eta_{j} - \lambda_{\hat{P}_{i}}
\eta_{\hat{P}_{i}}\right) = \vev{\hat{1} \hat{P}_{i}}^4 \,
\delta^{(4)}\left(\sum_{j=2}^{i-1} \frac{\vev{\hat{1}
{j}}}{\vev{\hat{1}\hat{P}_{i} }} \eta_{j} - \eta_{\hat{P}_{i}}
\right)\, \delta^{(4)}\left(\eta_{1} + \sum_{j=2}^{i-1}
\frac{\vev{j \hat{P}_{i}}}{\vev{\hat{1}\hat{P}_{i} }} \eta_{j}
\right)\,,
\ee
the integration over $\eta_{\hat{P}_{i}}$ can be
carried out straightforwardly. In this way, we obtain the
following contribution to the $n$-point NMHV amplitude:
\be
\label{B-result}
B = \frac{ \delta^{(4)}(p) \,
\delta^{(8)}(q)}{\prod_{j=1}^{n} \langle j \, j+1 \rangle } \,
\sum_{i=4}^{n-1} R_{n;2\,i} \,.
\ee
Here $R_{r;st}$ is
a dual superconformal invariant introduced in \cite{dhks5},
\be
\label{def-Rrst} R_{r;st} = \frac{\l<s
\,\, s-1\r> \l<t \,\, t-1\r>
\delta^{(4)}(\Xi_{r;st})}{x_{st}^2\l<r|x_{rs}x_{st}|t\r>\l<r|x_{rs}x_{st}|t-1\r>\l<r|x_{rt}x_{ts}|s\r>\l<r|x_{rt}x_{ts}|s-1\r>}\,.
\ee
The Grassmann odd quantity $\Xi_{r;st}$ is given by \be\label{defxi}
\Xi_{r;st} = \l<r|x_{rs}x_{st}|\theta_{tr}\r> +
\l<r|x_{rt}x_{ts}|\theta_{sr}\r>\,. \ee Here we used the dual
variables $x_{i}$ and $\theta_{i}$ defined by (\ref{dualx}) and (\ref{dualth}).

In the following we will often deal with the quantity $\Xi_{n;st}$ for $1< s<t < n$.
It is instructive to switch from the dual $\theta_i$ in (\ref{defxi}) to the $\eta_i$,
\be
\label{Xi-eta1}
\Xi_{n;st} = \langle n| \left\lbrack x_{ns} x_{st} \sum_{i=t}^{n-1} |i\rangle \eta_{i}+ x_{nt} x_{ts} \sum_{i=s}^{n-1} |i\rangle \eta_{i} \right\rbrack \,,
\ee
to see that $\Xi_{n;st}$ is independent of $\eta_{n}$ and $\eta_{1}$. Alternatively, using
the $\delta^{(8)}(q)$ present in all physical amplitudes to rewrite the sums we can obtain
\be
\label{Xi-eta2}
\delta^{(8)}(q)\,\Xi_{n;st} =- \delta^{(8)}(q)\, \langle n| \left\lbrack x_{ns} x_{st} \sum_{i=1}^{t-1} |i\rangle \eta_{i}+ x_{nt} x_{ts} \sum_{i=1}^{s-1} |i\rangle \eta_{i} \right\rbrack \,,
\ee
such that the only dependence on $\eta_{n-1}$ and $\eta_{n}$ on the l.h.s. of (\ref{Xi-eta2}) is contained in $\delta^{(8)}(q)$.
These facts will be useful in the following sections when carrying out superspace integrations.

Moreover, it is useful to realise that terms like $\l<r|x_{rs}x_{st}|t\r>$ in (\ref{def-Rrst}) and similar terms in (\ref{defxi}) can
always be written as
\be
\label{shift-aux}
\l<r|x_{rs}x_{st}|t\r> = \l<r|x_{r+1\,s}x_{st}|t\r>\,,
\ee
such that it is clear that they only depend explicitly on $\lambda_{r}$, but not on $\tilde{\lambda}_{r}$.

\subsection{5-point example}

In \cite{Brandhuber:2008pf}, the supersymmetric recursion relations were examined for the example of the five-point $\bMHV$ amplitude. We will also examine this example here as it is the first example of an NMHV amplitude.
For five points, NMHV${}_{5}$ = $\widebar{{\rm MHV}}_{5}$, and
therefore we could have obtained the NMHV${}_{5}$ amplitude from a
Grassmann Fourier transform of the ${\rm MHV}_{5}$ amplitude
\cite{dhks6}.

We immediately see that only the second term in
(\ref{super-BCF-NMHV}) contributes, because there is no four-point
amplitude of Grassmann degree $12$. Hence for five points, the
complete amplitude is given by (\ref{B-result}), i.e.
\be\label{NMHV5}
 {\cal A}^{\rm {NMHV}}_{5} = \frac{ \delta^{(4)}(p) \, \delta^{(8)}(q)}{\prod_{j=1}^{5} \langle j \, j+1 \rangle } \, R_{5;2\,4} \,.
\ee
We remark that the invariant $R_{5;2,4}$ can be further simplified, but this is a special feature of the $n=5$ case.

Another remark is that the super-amplitude must have cyclic
symmetry. This allows us to conclude that
\be \label{cyclic-5pt}
\delta^{(8)}(q)\, R_{5;2\,4}  = \delta^{(8)}(q)
R_{1;3\,5} = \delta^{(8)}(q)\, R_{2;4\,1} =
\delta^{(8)}(q)\, R_{3;5\,2} = \delta^{(8)}(q)\,
R_{4;1\,3}\,.
\ee
This is just the first
example of the more general identity for $n$ points, given in \cite{dhks6}, where
\be
\label{cyclic-npt} \delta^{(8)}(q)\, \sum_{s,t} R_{r;st} =
\delta^{(8)}(q)\, \sum_{s,t} R_{r';st}\,,
\ee
where the sum
goes over all values of $s,t$ such that $r,s,t$ (or $r',s,t$) are ordered cyclically with $r$ and $s$ (or $r'$ and $s$) and $s$ and $t$ separated by at least two.

\subsection{General solution for NMHV amplitudes}\label{NMHViteration}

It can be seen that there is a simple pattern to how the $n$-point solution is generated from the $(n-1)$-point one.
Let us check that the formula
\be\label{ansatzNMHV}
 {\cal A}^{\rm {NMHV}}_{n} = \mathcal{A}_n^{\rm MHV} \mathcal{P}_n^{\rm NMHV} =\frac{ \delta^{(4)}(p) \, \delta^{(8)}(q)}{\vev{1\, 2}\vev{2\, 3}\ldots\vev{n\, 1}} \, \sum_{2\leq s<t  \leq n-1} \!\!\!\!\!\! R_{n;st} \,,
\ee
indeed solves the supersymmetric recursion relation (\ref{NMHViteration}). In this formula we are assuming that $s$ and $t$ are separated by at least two. Comparing to (\ref{NMHV5}) we see that for $n=5$ the form (\ref{ansatzNMHV})
is correct.

We now proceed to prove (\ref{ansatzNMHV}) by induction.
Let us assume that the form (\ref{ansatzNMHV}) is valid for $n-1$ points.
Then it follows from the cyclicity of super-amplitudes that (\ref{cyclic-npt}) is also true for $n-1$ points.
Now, we notice that ${\cal A}^{\rm {NMHV}}_{n-1}(z_{P})$ in the homogeneous term, $A$ on the RHS of (\ref{super-BCF-NMHV}), only involves the quantities $R_{n-1;st}$ where
the first subscript is always equal to $n-1$.
Cyclic symmetry allows us
to insert ${\cal A}^{\rm {NMHV}}_{n-1}(z_{P})$ into (\ref{super-BCF-NMHV}) in our favourite
orientation. It is convenient to insert it such that the legs $\{ 1,2,3,\ldots,n-1\}$ of
${\cal A}^{\rm {NMHV}}_{n-1}(z_{P})$ are identified with the legs $\{ \hat{P},3,4,\ldots,n\}$ in the recursion
relation (see Fig. \ref{fig1}),
\be
A = \int
\frac{d^{4}P}{P^2} \int d^{4} \eta_{\hat{P}} \,
\mathcal{A}_{3}^{\rm \widebar{MHV}}(z_{P}) \mathcal{A}_{n-1}^{\rm
MHV} \mathcal{P}_{n-1}(\hat{P},3,\ldots,\bar{n}).
\ee
After carrying out this change of labels in ${\cal A}^{\rm {NMHV}}_{n-1}(z_{P})$ is is clear from equations (\ref{Xi-eta1}) and (\ref{shift-aux})
that the obtained $R_{n;st}$ does not depend on $\eta_{\hat{P}}$. Indeed the range of $\eta$-dependence is only $\{\eta_3,\ldots \eta_{n-1}\}$. When the lower summation variable attains its minimum value, there is an explicit dependence on the spinor $\l<\hat P|$. However, due to the three-point kinematics, this spinor is proportional to $\l< 2|$ and since it appears homogeneously in $R$ with degree zero it can simply be replaced by $\l<2|$. Thus we find
\be\label{induction-A} A = \frac{
\delta^{(4)}(p) \, \delta^{(8)}(q)}{\prod_{j=1}^{n} \langle j \,
j+1 \rangle } \, \sum_{3 \leq s<t \leq n-1} \!\!\!\!\!\! R_{n;st} \,. \ee
We see
that (\ref{B-result}) is just the missing first term (for $s=2$)
to complete (\ref{induction-A}) to the ansatz (\ref{ansatzNMHV})
for $n$ points, i.e.
\be A+B = {\cal A}^{\rm {NMHV}}_{n} = \frac{
\delta^{(4)}(p) \, \delta^{(8)}(q)}{\prod_{j=1}^{n} \langle j \,
j+1 \rangle } \, \sum_{2\leq s< t \leq n-1} \!\!\!\!\!\! R_{n;st} \,.
\ee
This
completes the inductive proof. Cyclicity of the super-amplitude
justifies the general identity (\ref{cyclic-npt}).
To prepare for the notation that we use in section \ref{sect-all}, we will rewrite the formula for NMHV amplitudes with different labels and using $\mathcal{P}_n^{\rm NMHV}$ instead of $\mathcal{A}_n^{\rm NMHV} = \mathcal{A}_n^{\rm MHV} \mathcal{P}_n^{\rm NMHV}$,
\be
\label{PNMHV}
\mathcal{P}_n^{\rm NMHV} = \sum_{2 \leq a_1,b_1 \leq n-1} \!\!\!\!\!\!\!\! R_{n;a_1b_1}.
\ee
The reason is that in the following sections we will derive a formula for the full $\mathcal{P}_n$ defined in (\ref{def-curlyP}) and we will encounter generalisations of the invariant $R_{n;a_1b_1}$ with multiple labels.

Thus we see that the result (\ref{ansatzNMHV}) which was conjectured in \cite{dhks5} and derived in \cite{dhks6} follows very naturally from the recursion relations. Of course it should be equivalent to the result found in \cite{Georgiou:2004by} using a supersymmetrised version of the CSW rules \cite{Cachazo:2004kj}.


\section{NNMHV tree amplitudes}\label{sect-NNMHV}

Before we generalise to all tree-level super-amplitudes, it is useful
to look first at the next case, namely NNMHV amplitudes. In examining
the recursion relation in this case we will find new features which
will help us find the solution for the full super-amplitude in the
next section.

The recursive relation for NNMHV amplitudes reads
\ba
\label{super-BCF-NNMHV}
\mathcal{A}_{n}^{\rm NNMHV} &=& \int
\frac{d^{4}P}{P^2} \int d^{4} \eta_{\hat{P}} \,
\mathcal{A}_{3}^{\rm \widebar{MHV}}(z_{P}) \mathcal{A}_{n-1}^{\rm
NNMHV}(z_{P}) + \sum_{i=4}^{n-3}  \int \frac{d^{4}P_{i}}{P_{i}^2}
\int d^{4}\eta_{\hat{P}_{i}} \mathcal{A}_{i}^{\rm MHV}(z_{P_{i}})
\mathcal{A}_{n-i+2}^{\rm NMHV}(z_{P_{i}}) \nonumber \\ &+&
\sum_{i=5}^{n-1}  \int \frac{d^{4}P_{i}}{P_{i}^2} \int
d^{4}\eta_{\hat{P}_{i}} \mathcal{A}_{i}^{\rm NMHV}(z_{P_{i}})
\mathcal{A}_{n-i+2}^{\rm MHV}(z_{P_{i}}) \equiv A +B_1 + B_2\,.
\ea
It is very similar to
the recursion relation for NMHV amplitudes, and as we will show
presently, it can be solved in a similarly straightforward manner.

\begin{figure}
\psfrag{dots}[cc][cc]{$\ldots$}
\psfrag{one}[cc][cc]{$\hat{1}$}
\psfrag{two}[cc][cc]{$2$}
\psfrag{three}[cc][cc]{$3$}
\psfrag{en}[cc][cc]{$\bar{n}$}
\psfrag{bigen}[cc][cc]{$NN$}
\psfrag{en2}[cc][cc]{$N$}
\psfrag{en3}[cc][cc]{$N$}
\psfrag{pe}[cc][cc]{$\hat{P}$}
\psfrag{pei}[cc][cc]{$\hat{P}_{i}$}
\psfrag{i}[cc][cc]{\hspace{0.5cm}$i-1$}
\psfrag{iplus}[cc][cc]{$i$}
\psfrag{A}[cc][cc]{$A$}
\psfrag{B}[cc][cc]{$B_{1}$}
\psfrag{C}[cc][cc]{$B_{2}$}
\psfrag{sumi}[cc][cc]{$\sum_{i=4}^{n-3}$}
\psfrag{sumj}[cc][cc]{$\sum_{i=5}^{n-1}$}
 \centerline{{\epsfysize2.8cm
\epsfbox{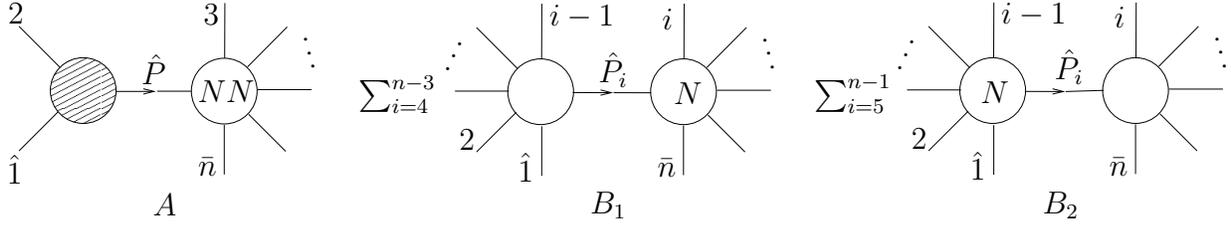}}} \caption[]{\small
  The three contributions to the supersymmetric recursion relation for
  NNMHV amplitudes.}
  \label{fig2}
\end{figure}

Before we derive the solution to (\ref{super-BCF-NNMHV}), it is helpful to introduce some new notation. Firstly we will introduce generalisations of the $R$-invariant which we used to express the NMHV amplitudes. The new quantities have many pairs of labels and are given by
\be
R_{n;b_1a_1;b_2a_2;\ldots;b_ra_r;ab} = \frac{\l< a\,\,a-1\r> \l<
  b\,\,b-1\r> \delta^{(4)}(\l< \xi | x_{a_r a}x_{ab} | \theta_{ba_r} \r> +
  \l< \xi |x_{a_r b} x_{ba} | \theta_{aa_r} \r>)}{x_{ab}^2 \l< \xi |
  x_{a_r a} x_{ab} | b \r> \l< \xi | x_{a_r a} x_{ab} |b-1\r> \l< \xi
  | x_{a_r b} x_{ba} |a\r> \l< \xi | x_{a_r b} x_{ba} |a-1\r>}\,,
  \label{generalR}
\ee
where the chiral spinor $\langle \xi |$ is given by
\be
\l< \xi | = \l<n | x_{nb_1}x_{b_1 a_1} x_{a_1 b_2} x_{b_2 a_2} \ldots x_{b_r a_r} \,.
\ee
In the case where there is only one pair of labels $ab$ after the initial label $n$, (\ref{generalR}) is just the $R$-invariant (\ref{def-Rrst}) we have already seen appearing in the NMHV amplitudes. The cases where there is more than one pair are generalisations.
The new quantities $R_{n;b_1a_1;\ldots;b_ra_r;ab}$ are invariant under dual conformal symmetry, but not (except for the case $R_{n;ab}$) under dual superconformal symmetry. However they will always appear in the amplitude together with additional factors which will combine with them to make dual superconformal invariants. We will explore this point in more detail in section \ref{sect-app-symmetry}.

We also need to introduce a second piece of notation. Just as we have already seen in the NMHV case, the $R$-invariants will always appear in the amplitude with a summation over the last pair of labels (the summation will always take place over the region where $a$ and $b$ are separated by at least two, $a<b-1$), i.e. in the form,
\be
\sum_{L\leq a<b \leq U} \!\!\!\!\!\! R_{n;b_1a_1;\ldots;b_ra_r;ab} \, .
\ee
We will write superscripts on the $R$-invariants to indicate special behaviour for the boundary terms when $a=L$ or $b=U$.
Specifically we write
\be
\sum_{L\leq a<b \leq U} \!\!\!\!\!\! R_{n;b_1a_1;\ldots;b_ra_r;ab}^{l_1\ldots l_p;u_1\ldots u_q} \, .
\label{superscripts}
\ee
This notation means the following. For the terms in the sum where $a=L$ we replace the explicit dependence on $\langle L-1|$ in (\ref{generalR}) in the following way,
\be
\langle L-1 | \longrightarrow \langle n|x_{nl_1} x_{l_1l_2} x_{l_2l_3} \ldots x_{l_{p-1} l_p} \, .
\label{Lrep}
\ee
Similarly, for the terms in the sum where $b=U$ we replace the explicit dependence on $\langle U|$ in (\ref{generalR}) in the following way,
\be
\langle U | \longrightarrow \langle n | x_{nu_1} x_{u_1u_2} x_{u_2u_3} \ldots x_{u_{q-1}u_q} \, .
\label{Urep}
\ee
Of course there is one term in the sum where $a=L$ and $b=U$ where both replacements occur. When no replacement is to be made on one of the boundaries we will write the superscript $0$.

Using this notation we will now state the result for all NNMHV amplitudes. As usual we have
\be
\mathcal{A}_n^{\rm NNMHV} = \mathcal{A}_n^{\rm MHV} \mathcal{P}_n^{\rm NNMHV}.
\label{ANNMHV}
\ee
Then the factor $\mathcal{P}_n^{\rm NNMHV}$ is given by
\begin{align}
\label{PNNMHVnew}
\mathcal{P}_n^{\rm NNMHV} = \sum_{2\leq a_1,b_1 \leq n-1}
\!\!\!\!\!\!\!\! R_{n;a_1b_1}^{0;0} \Bigl[ &\sum_{a_1+1 \leq a_2,b_2 \leq
  b_1} \!\!\!\!\!\!\!\! R_{n;b_1a_1;a_2b_2}^{0;a_1b_1}
  +  \sum_{b_1 \leq a_2b_2 \leq n-1} \!\!\!\!\!\!\!\! R_{n;a_2b_2}^{a_1b_1;0} \Bigr]\,.
\end{align}
The superscripts $0;0$ on the outer $R$-invariant $R_{n;a_1,b_1}^{0;0}$ simply mean that nothing special happens at the boundaries 2 and $n-1$, as is also the case in formula (\ref{PNMHV}) for the NMHV amplitudes. Thus this expression differs from the formula for the NMHV amplitudes in that the factor in the square brackets is not 1 but is itself a sum over $R$-invariants. For the sums of $R$-invariants in the square brackets the superscripts denote the fact that there are non-trivial boundary effects (at the upper boundary for the first term and the lower boundary for the second).

Let us now demonstrate the validity of formula (\ref{PNNMHVnew}). The first step is to calculate the two inhomogeneous terms in the recursion relation, labelled $B_1$ and $B_2$ in Fig. \ref{fig2}. We start with the calculation of $B_1$ which corresponds to the second term on the RHS of the recursion relation (\ref{super-BCF-NNMHV}). This term is very similar to the inhomogeneous term $B$ that we encountered for the NMHV amplitudes in section \ref{sect-NMHV}. The difference from that case is that for $B_1$ we have an additional factor of $\mathcal{P}^{\rm NMHV}$,
\be
B_1= \sum_{i=4}^{n-3}  \int \frac{d^{4}P_{i}}{P_{i}^2}
\int d^{4}\eta_{\hat{P}_{i}} \mathcal{A}_{i}^{\rm MHV}(z_{P_{i}})
\mathcal{A}_{n-i+2}^{\rm MHV}(z_{P_{i}}) \mathcal{P}_{n-i+2}^{\rm NMHV}(z_{P_{i}}) \, .
\ee
Thanks to the cyclic symmetry of the amplitudes, we have the freedom to insert the NMHV factor in our preferred orientation. We will choose to insert it so that the legs $\{ 1,2,\ldots,n-i+2 \}$ of the subamplitude correspond to the legs $\{ \hat{P},i,\ldots ,\bar{n} \}$ in the recursion relation, as shown in Fig. \ref{fig2}. With this choice we find that the $R$-invariants appearing in the factor of $\mathcal{P}^{\rm NMHV}_{n-i+2}$ (see equations (\ref{PNMHV}) and (\ref{def-Rrst})) do not depend on $\eta_{\hat{P}}$ and are therefore inert under the Grassmann integral. The integration is therefore identical to that which we performed in the calculation of $B$ in subsection \ref{subsect-inhom} and we obtain a result very similar to equation (\ref{B-result}),
\be
\label{B1-inter}
B_1 = \frac{ \delta^{(4)}(p) \,
\delta^{(8)}(q)}{\prod_{j=1}^{n} \langle j \, j+1 \rangle } \,
\sum_{i=4}^{n-1} R_{n;2\,i} \mathcal{P}_{n-i+2}^{\rm NMHV}(\hat{P},\ldots,\bar{n})\,.
\ee
Now, if we compare the factor $\mathcal{P}_{n-i+2}^{\rm NMHV}(\hat{P},\ldots,\bar{n})$ against the general formula for NMHV amplitudes (\ref{PNMHV}) and the definition of the $R$-invariants (\ref{def-Rrst}), we see that we can write it as
\be
\mathcal{P}^{\rm NMHV}_{n-i+2}(\hat{P},\ldots,\bar{n}) = \sum_{i \leq s,t \leq \bar{n}-1} \!\!\!\!\!\! R_{\bar{n};st}(\hat{P},\ldots,\bar{n}),
\label{Phatsum}
\ee
where the notation indicates that we must remember that legs associated to this factor form the ordered set $\{ \hat{P},\ldots,\bar{n} \}$. Thus when $s=i$ the explicit dependence of $R$ on $\langle s-1|$ becomes a dependence on $\langle \hat{P} |$. The spinor $\langle \hat{P} |$ appears once in the numerator and once in the denominator of the relevant $R$-invariants. For these boundary terms in the sum we will write the dependence on $\hat{P}$ in the following way. First we multiply both the numerator and denominator by $\l< n 1 \r>[1\hat{P}]$. Then we can see that for any factor which has the spinor $\langle \hat{P}|$ in it we can write
\begin{align}
\langle n 1 \rangle [1 \hat{P}] \langle \hat{P} |...  = \langle n 1 \rangle [1|P... = \langle n 1 \rangle [1 | x_{1i} ... = \langle n 1 \rangle [1|x_{2i} ... = \langle n |x_{12} x_{2i} ... = \langle n|x_{n2}x_{2i} ...
\label{removePhat}
\end{align}
So for the boundary terms $s=i$ we have a modification of the $R$-invariant where the explicit dependence on the spinor $\langle i-1|$ is replaced in the following way,
\be
\langle i-1| \longrightarrow \langle n|x_{n2}x_{2i} \, .
\label{Phatrep}
\ee
The remaining terms in the sum (\ref{Phatsum}) are just unmodified $R$-invariants. This is why we introduced the idea of superscripts on the $R$-invariants. The replacement (\ref{Phatrep}) is an example of the lower limit replacement (\ref{Lrep}). The total effect is summarised by the following formula for $B_1$,
\be
B_1 = \frac{ \delta^{(4)}(p) \, \delta^{(8)}(q)}{\prod_{j=1}^{n} \langle j \, j+1 \rangle } \, \sum_{i=4}^{n-1} R_{n;2\,i} \sum_{i \leq s,t \leq n-1} \!\!\!\!\!\! R_{n;st}^{2i;0}\, .
\ee
Now let us address the second inhomogeneous term $B_2$. This is similar to the term $B_1$ which we already calculated, but this time the factor of $\mathcal{P}_i^{\rm NMHV}$ appears in the left factor instead of the right factor,
\be
B_2= \sum_{i=4}^{n-3}  \int \frac{d^{4}P_{i}}{P_{i}^2}
\int d^{4}\eta_{\hat{P}_{i}} \mathcal{A}_{i}^{\rm MHV}(z_{P_{i}}) \mathcal{P}_i^{\rm NMHV}(z_{P_{i}})
\mathcal{A}_{n-i+2}^{\rm MHV}(z_{P_{i}}).
\ee
Again we can choose the legs of the left subamplitude so that the $R$-invariants contained in $\mathcal{P}_i^{\rm NMHV}$ are inert under the Grassmann integration. One way to do this is to have the legs $\{1,\ldots,i \}$ match up with legs $\{ 2,\ldots,-\hat{P},\hat{1} \}$ in the recursion relation. In much the same way as for $B_1$ this allows us to write
\be
\label{B2-inter}
B_2 = \frac{ \delta^{(4)}(p) \,
\delta^{(8)}(q)}{\prod_{j=1}^{n} \langle j \, j+1 \rangle } \,
\sum_{i=4}^{n-1} R_{n;2\,i} \sum_{3\leq s,t \leq \hat{P}} \!\!\!\! R_{\hat{1};st} (2,\ldots,-\hat{P},\hat{1})\,,
\ee
 where again the notation is to remind us that the legs associated with the $R$-invariants under the second sum form the ordered set $\{2,\ldots,-\hat{P},\hat{1} \}$. Thus when $t=\hat{P}$ we will have an explicit dependence on the spinor $\hat{P}$ in the $R$-invariants. Using exactly the same reasoning as in (\ref{removePhat}) above we see that the resulting $R$-invariants will have the upper boundary replacement,
\be
\langle i| \longrightarrow \langle n |x_{n2}x_{2i} \,.
\label{upperlimitrep}
\ee
In addition, there is a new feature in the calculation of $B_2$. This arises from the fact that the last leg in the subamplitude is $\hat{1}$ and not $n$. Therefore the spinor $\langle \hat{1} |$ appears four times in the numerator and four times in the denominator of every $R$-invariant. We can deal with this by writing the explicit expression for $\langle \hat{1}|$,
\be
\langle \hat{1} |  = \langle 1 | - z_{P_i} \langle n| = \langle 1| - \frac{x_{1i}^2}{\langle n|x_{1i}|1]}\langle n| = \frac{\langle n | x_{1i}(x_{12}  - x_{1i})}{\langle n |x_{1i} | 1]}=\frac{\langle n | x_{ni}x_{i2}}{\langle n | x_{1i} | 1]}.
\label{remove1hat}
\ee
Since $\langle \hat{1} |$ appears homogeneously in the $R$-invariants, the denominator $\langle n | x_{1i }| 1]$ in (\ref{remove1hat}) drops out and we effectively have the following replacement in the $R$-invariants,
\be
\langle n | \longrightarrow \langle n | x_{ni}x_{i2}\,.
\label{extension}
\ee
Taking into account both effects (\ref{upperlimitrep}) and (\ref{extension}) we find that $B_2$ is given by the following formula,
\be
B_2 = \frac{ \delta^{(4)}(p) \,
\delta^{(8)}(q)}{\prod_{j=1}^{n} \langle j \, j+1 \rangle } \,
\sum_{i=4}^{n-1} R_{n;2\,i} \sum_{3\leq s,t \leq i} \!\!\!\! R_{n;i2;st}^{0;2i} \, .
\ee
The upper limit replacement (\ref{upperlimitrep}) is responsible for the non-trivial right-superscript, while the extension of the spinor $\langle n |$ in (\ref{extension}) is responsible for the fact that we have the first example of the generalised $R$-invariants, defined in equation (\ref{generalR}).

Now we are in a position to justify the formula ({\ref{PNNMHVnew}) for the NNMHV amplitudes. We will proceed by induction and assume that (\ref{PNNMHVnew}) is true for $(n-1)$-point amplitudes. Then we can treat the homogeneous term (labelled $A$ in Fig. \ref{fig2}) in exactly the same way as for NMHV amplitudes. Again we will insert $\mathcal{A}^{\rm NNMHV}_{n-1}(z_P)$ so that the legs $\{ 1,\ldots,n-1 \}$ of the subamplitude coincide with legs $\{ \hat{P},3,\ldots,\bar{n} \}$ of the recursion relation, 
\be
A = \int
\frac{d^{4}P}{P^2} \int d^{4} \eta_{\hat{P}} \,
\mathcal{A}_{3}^{\rm \widebar{MHV}}(z_{P}) \mathcal{A}_{n-1}^{\rm
MHV}(z_{P}) \mathcal{P}_{n-1}(\hat{P},3,\ldots,\bar{n}).
\ee
With this choice we find all $R$-invariants are again inert under the Grassmann integral. To see this, we first note that the outer $R$-invariant in $\mathcal{P}_{n-1}(\hat{P},3,\ldots,\bar{n})$ (see (\ref{PNNMHVnew})) is the same as in the NMHV case. We have already seen in subsection \ref{NMHViteration} that this does not depend on $\eta_{\hat{P}}$ and so is inert under the Grassmann integral. The other $R$-invariants in $\mathcal{P}_{n-1}(\hat{P},3,\ldots,\bar{n})$ (which come from the terms in square brackets in (\ref{PNNMHVnew})) also do not depend on $\eta_{\hat{P}}$. The first term in the square brackets depends on $\{\eta_3,\ldots,\eta_{n-2} \}$, as can be seen from equations (\ref{generalR}) and (\ref{dualth}), while the second depends on $\{ \eta_{3},\ldots,\eta_{n-1} \}$ just like the outer $R$-invariant.

Just as for the case of the NMHV amplitudes, when the outermost lower summation variable (which corresponds to $a_1$ in equation (\ref{PNNMHVnew})) reaches its lowest value, there is an explicit dependence on the spinor $\langle \hat{P}|$. However, as in the NMHV case, this can simply be replaced by $\langle 2 |$ due to the three-point kinematics. Thus we obtain the following simple result for $A$,
\be
A =  \frac{ \delta^{(4)}(p) \, \delta^{(8)}(q)}{\prod_{j=1}^{n} \langle j \, j+1 \rangle } \, \sum_{3\leq a_1,b_1 \leq n-1}
\!\!\!\!\!\!\!\! R_{n;a_1b_1}^{0;0} \Bigl[ \sum_{a_1+1 \leq a_2,b_2 \leq
 b_1} \!\!\!\!\!\!\!\! R_{n;b_1a_1;a_2b_2}^{0;a_1b_1}
 +  \sum_{b_1 \leq a_2b_2 \leq n-1} \!\!\!\!\!\!\!\! R_{n;a_2b_2}^{a_1b_1;0} \Bigr]\,.
\ee
Combining the results from $A$, $B_1$ and $B_2$ we find
\begin{align}
A + B_1 + B_2  &= \frac{ \delta^{(4)}(p) \, \delta^{(8)}(q)}{\prod_{j=1}^{n} \langle j \, j+1 \rangle } \, \sum_{2\leq a_1,b_1 \leq n-1}
\!\!\!\!\!\!\!\! R_{n;a_1b_1}^{0;0} \Bigl[ \sum_{a_1+1 \leq a_2,b_2 \leq
 b_1} \!\!\!\!\!\!\!\! R_{n;b_1a_1;a_2b_2}^{0;a_1b_1}
 +  \sum_{b_1 \leq a_2b_2 \leq n-1} \!\!\!\!\!\!\!\! R_{n;a_2b_2}^{a_1b_1;0} \Bigr]\, \notag\\
&= \mathcal{A}^{\rm MHV}_n \mathcal{P}^{\rm NNMHV}_n.
\end{align}
We know formula (\ref{PNNMHVnew}) is correct for the six-point amplitudes, since the inhomogeneous terms are the only contributions to this case. Therefore we have completed the inductive justification of the result (\ref{PNNMHVnew}) for the NNMHV amplitudes.

\section{All tree amplitudes}\label{sect-all}

It is simple to continue the analysis of the preceding sections to
N${}^3$MHV, N${}^4$MHV amplitudes and so on. The supersymmetric
recursion relation for a generic N${}^{p}$MHV
amplitude reads \ba \label{super-BCF-all} \mathcal{A}_{n}^{{\rm
N}^{p}{\rm MHV}} &=& \int \frac{d^{4}P}{P^2} \int d^{4}
\eta_{\hat{P}} \, \mathcal{A}_{3}^{\rm \widebar{MHV}}(z_{P})
\mathcal{A}_{n-1}^{{\rm N}^{p}{\rm MHV}}(z_{P}) \nonumber \\
 &+& \sum_{m=0}^{p-1} \; \; \sum_i  \int \frac{d^{4}P_{i}}{P_{i}^2}
\int d^{4}\eta_{\hat{P}_{i}}
 \mathcal{A}_{i}^{{\rm N}^{m}{\rm MHV}}(z_{P_{i}})
 \mathcal{A}_{n-i+2}^{{\rm N}^{(p-m-1)}{\rm MHV}}(z_{P_{i}})\,.
\ea
At each stage one obtains the universal prefactor $\mathcal{A}_n^{\rm
  MHV}$ while the $R$-invariants from the right-hand factor in the
second line are left unchanged and those from the left-hand factor
acquire an additional extension, just as in the case of the NNMHV
amplitudes. As before, one must carefully take into account the behaviour of the boundary terms in the sums.
For example, we find that the N${}^3$MHV amplitudes are given by the formula,
\begin{align}\label{NNNMHV-1}
\mathcal{P}^{\rm{N}^3\rm{MHV}}_n = \sum_{2\leq a_1,b_1 \leq n-1}
\!\!\!\!\!\! R_{n;a_1b_1}
\Bigl[&\sum_{a_1+1 \leq a_2,b_2 \leq b_1} \!\!\!\!\!\! R_{n;b_1a_1;a_2b_2}^{0;a_1b_1}
  \Bigl(\sum_{a_1+1 \leq a_3,b_3 \leq b_2}  \!\!\!\!\!\! R_{n;b_1a_1;b_2a_2;a_3b_3}^{0;b_1a_1a_2b_2}+
  \sum_{b_2 \leq a_3,b_3 \leq b_1} \!\!\!\!  R_{n;b_1a_1;a_3b_3}^{b_1a_1a_2b_2;a_1b_1} \Bigr) \notag \\
+&\sum_{a_1+1 \leq a_2,b_2 \leq b_1} \!\!\!\!\!\! R_{n;b_1a_1;a_2b_2}^{0;a_1b_1} \sum_{b_1 \leq
  a_3,b_3 \leq n-1} \!\!\!\!\!\! R_{n;a_3b_3}^{a_1b_1;0} \notag \\
+&\sum_{b_1\leq a_2,b_2 \leq n-1} \!\!\!\!\!\! R_{n;a_2b_2}^{a_1b_1;0}\Bigl(\sum_{a_2+1\leq
  a_3,b_3 \leq b_2} \!\!\!\!\!\! R_{n;b_2a_2;a_3b_3}^{0;a_2b_2} + \sum_{b_2 \leq a_3,b_3 \leq n-1}
\!\!\!\!\!\! R_{n;a_3b_3}^{a_2b_2;0}\Bigr)
\Bigr] \,.
\end{align}
If we take the terms in the outermost sum where $a_1=2$ then the three lines correspond to the three different inhomogeneous terms in the recursion relation $\mathcal{A}^{\rm NNMHV}_L \mathcal{A}^{\rm
  MHV}_R$, $\mathcal{A}^{\rm NMHV}_L \mathcal{A}^{\rm NMHV}_R$ and
$\mathcal{A}^{\rm MHV}_L \mathcal{A}^{\rm NNMHV}_R$. As before, the superscripts on the $R$-invariants indicate the lower and upper limit replacements. The formula (\ref{NNNMHV-1}) can be justified by induction, just as we saw in the cases of the NMHV and NNMHV amplitudes. We will not give the argument here because in this section we will give an inductive argument which proves a general formula for the whole super-amplitude (i.e. for all N${}^p$MHV amplitudes for all $p$).

It is helpful to notice that the first and second lines of (\ref{NNNMHV-1}) can be combined so that we have
\begin{align}\label{NNNMHV-2}
&\mathcal{P}^{\rm{N}^3\rm{MHV}}_n = \sum_{2\leq a_1,b_1 \leq n-1}
  \!\!\!\!\!\!\!R_{n;a_1b_1}\biggl[\notag \\
&\sum_{a_1+1 \leq a_2,b_2 \leq b_1} \!\!\!\!\!\!\!\!\!
    R_{n;b_1a_1;a_2b_2}^{0;a_1b_1} \Bigl(\sum_{a_2+1 \leq a_3,b_3 \leq b_2}
    \!\!\!\!\!\!\!\!\!R_{n;b_1a_1;b_2a_2;a_3b_3}^{0;b_1a_1a_2b_2} + \sum_{b_2 \leq
      a_3,b_3 \leq b_1} \!\!\!\!\!\!R_{n;b_1a_1;a_3b_3}^{b_1a_1a_2b_2;a_1b_1}+ \sum_{b_1
      \leq a_3,b_3 \leq n-1} \!\!\!\!\!\!\!\! R_{n;a_3b_3}^{a_1b_1;0}
    \Bigr)\notag \\
& +\sum_{b_1\leq a_2,b_2 \leq n-1} \!\!\!\!\!
    R_{n;a_2b_2}^{a_1b_1;0}\Bigl(\sum_{a_2+1\leq a_3,b_3 \leq b_2} \!\!\!\!\!\!
    R_{n;b_2a_2;a_3b_3}^{0;a_2b_2} + \sum_{b_2 \leq a_3,b_3 \leq n-1}
    \!\!\!\!\!\!R_{n;a_3b_3}^{a_2b_2;0}\Bigr)
\biggr] \,.
\end{align}
The reason we group the terms in this way is that it fits very naturally, together with formulae (\ref{PNMHV}) and (\ref{PNNMHVnew}) for the NMHV and NNMHV cases, into a general pattern which we will now describe.

In the remainder of this section we will prove a general formula for all tree-level amplitudes in $\mathcal{N}=4$ super Yang-Mills. First we must state the result. In order to do so we need to introduce a diagrammatic way of organising the general formula. Then we will go on to prove the formula by induction.

We illustrate the full
$n$-point super-amplitude in Fig. \ref{fig-rec-solution} as a tree diagram, where the vertices correspond to the different $R$-invariants which appear. We consider
a rooted tree, with the top vertex (the root) denoted by 1. The root
has a single descendant vertex with labels $a_1,b_1$ and
the tree is completed by passing from each vertex to a number of
descendant vertices, as described in Fig. \ref{fig-generic}. We will enumerate the rows by
$0,1,2,3,\ldots$ with 0
corresponding to the root.
For an
$n$-point super-amplitude (with $n\geq 4$) only the rows up to row $n-4$ in the
tree will contribute to the amplitude\footnote{The three-point MHV amplitude is a special case where
  only the root vertex contributes.}.
The rule for completing the tree as given
in Fig. \ref{fig-generic} can be easily seen to imply that the number
of vertices in row $p$ is the Catalan number $C(p) = (2p)! / (p!
(p+1)!)$.

\begin{figure}
\psfrag{one}[cc][cc]{$1$}
\psfrag{a1b1}[cc][cc]{$a_{1}b_{1}$}
\psfrag{a2b2}[cc][cc]{$a_{2}b_{2}$}
\psfrag{a3b3}[cc][cc]{$a_{3}b_{3}$}
\psfrag{b1a1a2b2}[cc][cc]{$b_{1}a_{1};a_{2}b_{2}$}
\psfrag{b2a2a3b3}[cc][cc]{$b_{2}a_{2};a_{3}b_{3}$}
\psfrag{b1a1a3b3}[cc][cc]{$b_{1}a_{1};a_{3}b_{3}$}
\psfrag{b1a1b2a2a3b3}[cc][cc]{$b_{1}a_{1};b_{2}a_{2};a_{3}b_{3}$}
\psfrag{two}[cc][cc]{$2$}
\psfrag{n1}[cc][cc]{$n-1$}
\psfrag{a1p}[cc][cc]{$a_{1}+1$}
\psfrag{a2p}[cc][cc]{$a_{2}+1$}
\psfrag{b1}[cc][cc]{$b_{1}$}
\psfrag{b2}[cc][cc]{$b_{2}$}
 \centerline{{\epsfysize7cm
\epsfbox{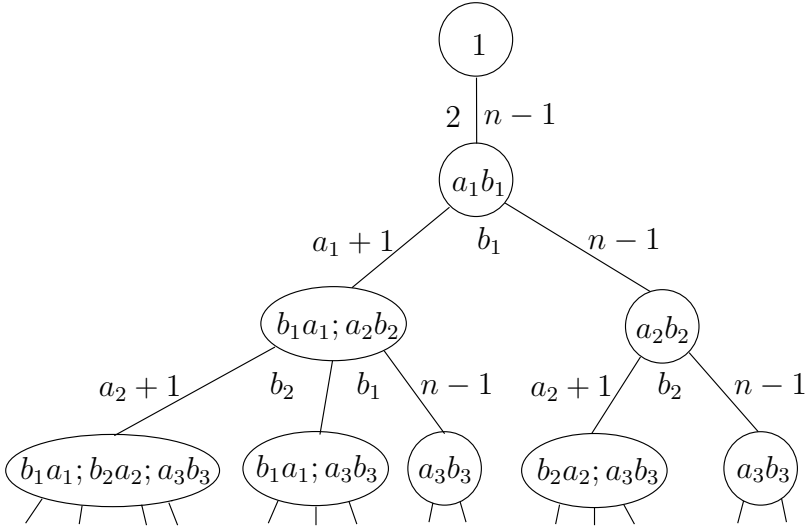}}}  \caption[]{\small
  Graphical representation of the formula for tree-level amplitudes in
  $\mathcal{N}=4$ SYM.}
  \label{fig-rec-solution}
\end{figure}

Each vertex in the tree corresponds to an $R$-invariant with first
label $n$ and the remaining labels corresponding to
those written in the vertex. For example, the first descendant vertex corresponds to the invariant $R_{n;a_1b_1}$ which we already saw appearing from the NMHV level. The next descendant vertices correspond to $R_{n;b_1a_1;a_2b_2}$ (which appears for the first time at NNMHV level) and $R_{n;a_2b_2}$, etc.

We consider vertical paths in the tree,
starting from the root vertex at the top of Fig. \ref{fig-rec-solution}. To each path we associate the product of the
$R$-invariants (vertices) visited by the path, with a nested summation
over all labels. The last pair of labels in a given vertex correspond
to the ones which are summed first, i.e. the ones of the inner-most
sum. In row $p$ they are denoted by $a_p,b_p$. We always take the convention that $a_p + 2 \leq b_p$, which is needed for the corresponding $R$-invariant to be well-defined.

The lower and upper limits for the summation over the pair of labels $a_p,b_p$ are noted to the left and right of the line above each vertex in row $p$. For example, the labels $a_1$ and $b_1$ of $R_{n;a_1,b_1}$, associated to the first descendant vertex, are to be summed over the region $2\leq a_1,b_1 \leq n-1$, as always with the convention that $a_1+2 \leq b_1$. The labels $a_2$ and $b_2$ on the $R$-invariants associated to the next descendant vertices are summed over the region $a_1+1\leq a_2,b_2 \leq b_1$ for the vertex on the left, and the region $b_1 \leq a_2,b_2 \leq n-1$ for the vertex on the right, in both cases with the condition $a_2 +2 \leq b_2$.

As we have seen already in the case of the NNMHV amplitudes, sometimes the $R$-invariants need to be modified when the summation labels reach their limiting lower or upper values. We deal with this by writing superscripts on the corresponding $R$-invariants, as we described in equations (\ref{superscripts},\ref{Lrep},\ref{Urep}). We will illustrate how to obtain the superscripts on each $R$-invariant by referring to a general cluster of vertices in row $p$ with a common parent vertex in row $p-1$, as shown in Fig. \ref{fig-generic}. Firstly, the left superscript of the left-most vertex in the cluster and the right superscript of the right-most vertex are both 0, i.e. they indicate no replacements at these boundaries. Then for the rest, the left superscript associated to a given vertex coincides with the right superscript associated to the vertex to its left. Therefore we need only specify the right superscripts. These are given by taking the labels in the vertex, deleting the
  final pair $a_p,b_p$ and then reversing the order of the last pair which remain. For example, the vertex second from the left in Fig. \ref{fig-generic} corresponds to the following sum,
\begin{align}
\sum_{b_{p-1} \leq a_pb_p \leq v_r} \!\!\!\!\!\! R_{n;v_1u_1;\ldots;v_ru_r;a_pb_p}^{v_1u_1\ldots v_ru_ra_{p-1}b_{p-1};v_1u_1\ldots v_{r-1}u_{r-1}u_rv_r}  \, .
\end{align}
The right superscript on the $R$-invariant is determined by taking the labels in the vertex, $v_1u_1,\ldots,v_ru_r,a_pb_p$, deleting the final pair $a_pb_p$, and then reversing the order of the final two which remain, namely $v_ru_r$. The left superscript coincides with the right superscript of the vertex to its left in Fig. \ref{fig-generic}, and so can be obtained by performing the same operation on the labels of that vertex.

The formula for
the full super-amplitude $\mathcal{A}_n = \mathcal{A}_n^{\rm MHV} \mathcal{P}_n$ is given by the sum over all vertical paths
of any length, starting from the root,
\be
\label{paths}
\mathcal{P}_n = \sum \text{vertical paths in Fig. \ref{fig-rec-solution}}.
\ee

\begin{figure}
\psfrag{uvs}[cc][cc]{$v_{1}u_{1};\ldots v_{r}u_{r};a_{p-1}b_{p-1}$}
\psfrag{uvab}[cc][cc]{$v_{1}u_{1};\ldots v_{r}u_{r};b_{p-1}a_{p-1};a_{p}b_{p}$}
\psfrag{uvnext}[cc][cc]{$v_{1}u_{1};\ldots v_{r}u_{r};a_{p}b_{p}$}
\psfrag{ab}[cc][cc]{$a_{p}b_{p}$}
\psfrag{up}[cc][cc]{$a_{p-1}+1$}
\psfrag{bp1}[cc][cc]{$b_{p-1}$}
\psfrag{bvr}[cc][cc]{$v_{r}$}
\psfrag{bv1}[cc][cc]{$v_{1}$}
\psfrag{n1}[cc][cc]{$n-1$}
\psfrag{dots}[cc][cc]{$\ldots$}
 \centerline{{\epsfysize4cm
\epsfbox{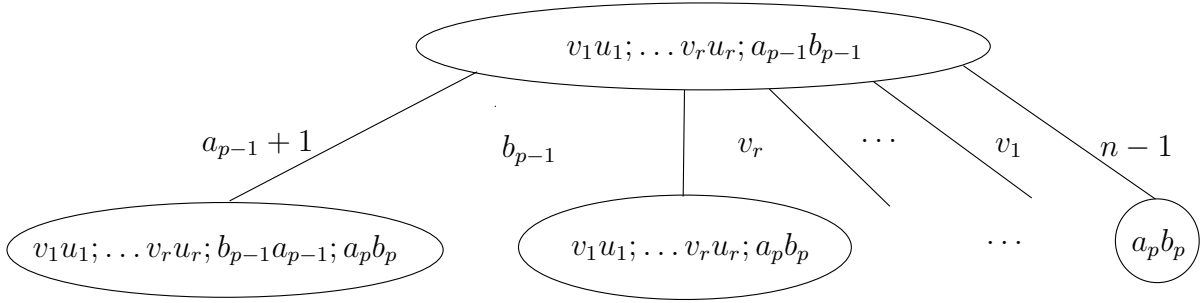}}}  \caption[]{\small
  The rule for going from line $p-1$ to line $p$ (for $p>1$) in Fig. \ref{fig-rec-solution}. For every vertex in line $p-1$ of the form given at the top of the diagram, there are $r+2$ vertices in the lower line (line $p$). The labels in these vertices start with $v_{1}u_{1};\ldots v_{r}u_{r};b_{p-1}a_{p-1};a_{p}b_{p}$ and they get sequentially shorter, with each step to the right removing the pair of labels adjacent to the last pair $a_p,b_p$ until only the last pair is left. The summation limits between each line are also derived from the labels of the vertex above.
The right superscripts associated to each vertex are obtained by deleting the final pair of labels $a_pb_p$ and reversing the order the last pair which remain. The left superscript of a given vertex coincides with the right superscript of the vertex to its left.
 }
  \label{fig-generic}
\end{figure}

Let us now see how the formula (\ref{paths}) works for the first few cases. Firstly there is one path of length zero, where we start at the root (row zero) and do not go anywhere. The value of this path is simply 1 and it corresponds to the MHV amplitudes,
\be
\mathcal{P}_n^{\rm MHV} = 1\,.
\ee
There is one path of length one, where we start at the root and go one step to its unique descendant. This path gives us $1$ from the root, multiplied by $R_{n;a_1b_1}$ from the descendant vertex, summed over $a_1,b_1$ with lower limit $2$ and upper limit $n-1$. There are no boundary replacements in the sum since there is only one $R$-invariant in the relevant cluster and so both its left and right superscripts are 0. So we obtain for the NMHV amplitudes,
\be
\mathcal{P}_n^{\rm NMHV} = \sum_{2 \leq a_1,b_1 \leq n-1} \!\!\!\!\!\!\!\! R_{n;a_1b_1}^{0;0} = \sum_{2 \leq a_1b_1 \leq n-1} \!\!\!\!\!\!\!\! R_{n;a_1,b_1}\,,
\ee
which agrees with eq. (\ref{PNMHV}).

There are two paths of length two. The first corresponds to descending from the root by one step and then descending once more to the left in Fig. \ref{fig-rec-solution}. For this path we obtain $1$ multiplied by $R_{n;a_1b_1}$ multiplied by $R_{n;b_1a_1;a_2b_2}$ with the limits for the outer sum over $a_1,b_1$ being the same as for the NMHV case above, while the inner sum, which is over $a_2,b_2$, has lower limit $a_1+1$ and upper limit $b_1$. The second path of length two corresponds to descending to the right instead of to the left. Doing so we obtain the product $1 \times R_{n;a_1b_1} \times R_{n;a_2b_2}$ with summation limits in the outer sum as before and in the inner sum being $b_1 \leq a_2,b_2 \leq n-1$.

The superscripts on the factors $R_{n;a_1b_1}$ are trivial as we just saw when looking at paths of length one.
To obtain the superscripts on the other $R$-invariants, we recall that the left superscript of the left-most vertex in row 2 of Fig. \ref{fig-rec-solution} (corresponding to $R_{n;b_1a_1;a_2b_2}$) and also the right superscript of the right-most vertex (corresponding to $R_{n;a_2b_2}$) are 0. There is one non-trivial right superscript, that of $R_{n;b_1a_1;a_2b_2}$. It is obtained by deleting the final pair of indices $a_2b_2$ and reversing the order of the last pair which remains (which in this case is the pair $b_1a_1$). Thus we obtain $R_{n;b_1a_1;a_2b_2}^{0;a_1b_1}$. The left superscript of the other invariant is the same and so we obtain $R_{n;a_2b_2}^{a_1b_1;0}$.

Adding the two paths we obtain for the NNMHV amplitudes
\be
\mathcal{P}_n^{\rm NNMHV} = \sum_{2 \leq a_1,b_1 \leq n-1} \!\!\!\!\!\!\!\! R_{n;a_1b_1}^{0;0} \Bigl( \sum_{a_1+1 \leq a_2,b_2 \leq b_1} \!\!\!\!\!\!\!\! R_{n;b_1a_1;a_2b_2}^{0;a_1b_1} + \sum_{b_1 \leq a_2,b_2 \leq n-1} \!\!\!\!\!\!\!\! R_{n;a_2b_2}^{a_1b_1;0} \Bigr)\,,
\ee
which agrees with eq. (\ref{PNNMHVnew}).

Continuing, we find five paths of length three. Applying the rules for writing the sums over $R$-invariants and specifying their superscripts we find they correspond precisely to the five terms in the expression (\ref{NNNMHV-2}) for the N${}^3$MHV amplitudes. Generically, since the number of vertices in row $p$ of the tree in Fig. \ref{fig-rec-solution} is the Catalan number $C(p)$, we find $C(p)$ terms in the expression for the N${}^p$MHV amplitudes. Finally, by considering the sum of all vertical paths of any length, starting from the root, we obtain the sum of all amplitudes,
\be
\mathcal{P}_n = \mathcal{P}_n^{\rm MHV} + \mathcal{P}_n^{\rm NMHV} + \mathcal{P}_n^{\rm NNMHV} + \ldots + \mathcal{P}_n^{\overline{\rm MHV}}.
\ee
The sum terminates (as it should) because, for a given value of $n$, there is maximum number of possible nestings beyond which all sums collapse to zero. This means that only paths up to length $n-4$ contribute and the longest paths correspond to the $\overline{\rm MHV}$ amplitudes.
This completes the statement of the result for all tree-level amplitudes.

We will now prove the validity of formula (\ref{paths}), i.e. that all tree amplitudes are indeed given by summing vertical paths in the tree diagram Fig. \ref{fig-rec-solution}. As usual we will proceed by induction and assume that the formula is correct for $(n-1)$-point amplitudes. The recursion relation for the full superamplitude $\mathcal{A}_n$ is illustrated in Fig. \ref{fig-fullrecur}. All vertices except the $\overline{\rm MHV}_3$ vertex are full super-amplitudes. Specifically the relation reads
\begin{align}
\mathcal{A}_n = \int \frac{d^4P}{P^2} \int d \eta_P \mathcal{A}^{\overline{\rm MHV}}_3(z_P) \mathcal{A}_{n-1}(z_P) + \sum_{i=4}^{n-1} \int \frac{d^4P_i}{P_i^2} \int d\eta_{P_i} \mathcal{A}_i (z_{P_i}) \mathcal{A}_{n-i+2}(z_{P_i}).
\label{fullrecursion}
\end{align}
We will call the first term on the RHS side the linear term because it is linear in the full super-amplitude $\mathcal{A}$. Similarly we call the second term the quadratic term because there are two factors of $\mathcal{A}$ for each term in the sum over $i$.

\begin{figure}
\psfrag{dots}[cc][cc]{$\ldots$}
\psfrag{one}[cc][cc]{$\hat{1}$}
\psfrag{two}[cc][cc]{$2$}
\psfrag{three}[cc][cc]{$3$}
\psfrag{en}[cc][cc]{$\bar{n}$}
\psfrag{bigen}[cc][cc]{$$}
\psfrag{pe}[cc][cc]{$\hat{P}$}
\psfrag{pei}[cc][cc]{$\hat{P}_{i}$}
\psfrag{i}[cc][cc]{\hspace{0.5cm}$i-1$}
\psfrag{iplus}[cc][cc]{$i$}
\psfrag{A}[cc][cc]{$$}
\psfrag{B}[cc][cc]{$$}
\psfrag{sumi}[cc][cc]{$\sum_{i=4}^{n-1}$}
 \centerline{{\epsfysize3cm
\epsfbox{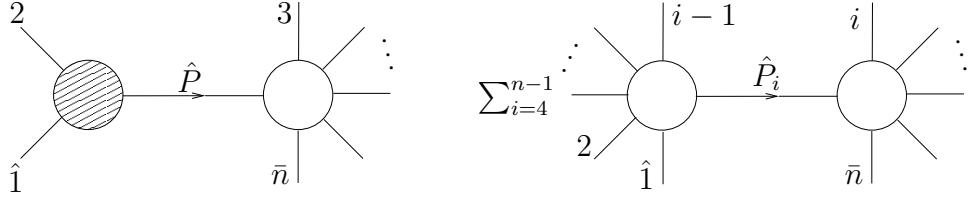}}}  \caption[]{\small
  The two contributions to the RHS of the supersymmetric recursion relation for the full super-amplitudes. We call the first term the linear term and the second term the quadratic term. As before
    $\hat{1}$ means that $\lambda_{1}$ is shifted, and $\bar{n}$ means that $\tilde{\lambda}_{n}$ is shifted.}
  \label{fig-fullrecur}
\end{figure}

We will introduce $\mathcal{P}_n$ into (\ref{fullrecursion}) in the usual way, $\mathcal{A}_n = \mathcal{A}_n^{\rm MHV} \mathcal{P}_n$.
As in the particular cases of NMHV and NNMHV amplitudes, it is useful to insert the subamplitudes in this expression in our favourite orientations. We will choose the same orientations that we chose in those cases, i.e. a left factor will depend on the ordered set $\{2,\ldots,-\hat{P},\hat{1} \}$ and a right factor on the ordered set $\{ \hat{P},\ldots,\bar{n} \}$. With this choice the recursion relation (\ref{fullrecursion}) reads
\begin{align}
\mathcal{A}_n^{\rm MHV} \mathcal{P}_n = &\int \frac{d^4P}{P^2} \int d \eta_P \mathcal{A}^{\overline{\rm MHV}}_3(z_P) \mathcal{A}_{n-1}^{\rm MHV} \mathcal{P}_{n-1}(\hat{P},3,\ldots,\bar{n}) \notag\\
&+ \sum_{i=4}^{n-1} \int \frac{d^4 P_i}{P_i^2} \int d\eta_{P_i} \mathcal{A}_i^{\rm MHV} \mathcal{P}_i (2,\ldots,-\hat{P_i},\hat{1}) \mathcal{A}_{n-i+2}^{\rm MHV} \mathcal{P}_{n-i+2}(\hat{P_i},i,\ldots,\bar{n}) .
\label{fullrecinter}
\end{align}
The reason for making this particular choice of orientations for the subamplitudes is the same as in the NMHV and NNMHV cases; the $\mathcal{P}$ factors are all inert under the Grassmann integral. This can be seen by looking at the $\eta$-dependence of the $R$-invariants appearing in the $\mathcal{P}$ factors, defined by the sum over paths in the tree diagram Fig. \ref{fig-rec-solution}. The outer most $R$-invariant in each $\mathcal{P}$ factor is the same as in $\mathcal{P}^{\rm NMHV}$ (which we have already seen is inert with this choice of orientation) and the other $R$-invariants have at least as restrictive a range of $\eta$-dependence. This is just as we saw in the the case of the NNMHV amplitudes. 

Once we have seen that the $\mathcal{P}$-factors are all inert, the Grassmann integrals in (\ref{fullrecinter}) are simple to do. The integration in the first term is the same as in the terms we called $A$ in the NMHV and NNMHV cases, it provides the usual $\mathcal{A}_n^{\rm MHV}$ factor and leaves the $\mathcal{P}$-factor unchanged. In the second term, the Grassmann integration is the same as in the term we called $B$ in the NMHV case or those we called $B_1$ and $B_2$ in the NNMHV case. We obtain a factor of $R_{n;2i}$ for each $i$ as well as a factor of $\mathcal{A}_n^{\rm MHV}$. Thus we have
\be
\mathcal{P}_n = \mathcal{P}_{n-1}(\hat{P},3,\ldots,\bar{n}) + \sum_{i=4}^{n-1} R_{n;2,i} \mathcal{P}_i(2,\ldots,-\hat{P_i},\hat{1}) \mathcal{P}_{n-i+2}(\hat{P_i},i,\ldots,\bar{n}).
\label{fullPrec}
\ee
As we saw already in the NMHV and NNMHV cases, the spinors $\langle \hat{P} |$ appearing in the $R$-invariants in the first term on the RHS (the linear term) can be replaced by $\langle 2|$ due to the three-point kinematics.
This term then gives an expression which is almost identical to the sum over paths in Fig. \ref{fig-rec-solution}, except that the lower limit of the outermost sum is 3 and not 2. Thus to prove the inductive step we need to show that the second term in (\ref{fullPrec}) (the quadratic term) gives the missing contributions, i.e. those paths of length one or greater where the outermost lower summation variable is fixed to be 2. 

In fact the contributions we are looking for can also be represented diagrammatically as a sum over paths in a tree diagram very similar to the tree in Fig. \ref{fig-rec-solution}. The tree representing the missing paths differs from Fig. \ref{fig-rec-solution} in that the root vertex is missing and the label $a_1$ is fixed to the value 2. We must remember that the label $b_1$ is still summed over the range $4\leq b_1 \leq n-1$. We give the relevant tree diagram in Fig. \ref{missingpathstree}.

\begin{figure}
\psfrag{a1b1}[cc][cc]{$2\, b_{1}$}
\psfrag{a2b2}[cc][cc]{$a_{2}b_{2}$}
\psfrag{a3b3}[cc][cc]{$a_{3}b_{3}$}
\psfrag{b1a1a2b2}[cc][cc]{$b_{1}\,2;a_{2}b_{2}$}
\psfrag{b2a2a3b3}[cc][cc]{$b_{2}a_{2};a_{3}b_{3}$}
\psfrag{b1a1a3b3}[cc][cc]{$b_{1}\,2;a_{3}b_{3}$}
\psfrag{b1a1b2a2a3b3}[cc][cc]{$b_{1}\,2;b_{2}a_{2};a_{3}b_{3}$}
\psfrag{n1}[cc][cc]{$n-1$}
\psfrag{a1p}[cc][cc]{$3$}
\psfrag{a2p}[cc][cc]{$a_{2}+1$}
\psfrag{b1}[cc][cc]{$b_{1}$}
\psfrag{b2}[cc][cc]{$b_{2}$}
 \centerline{{\epsfysize6cm
\epsfbox{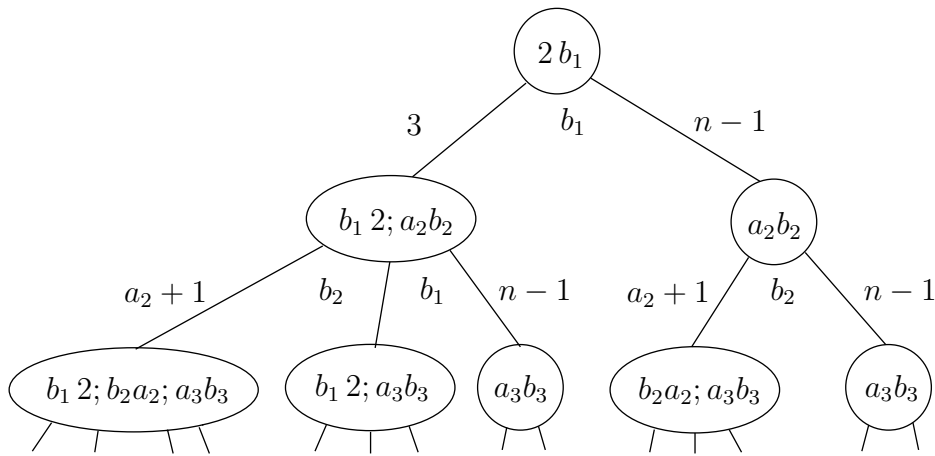}}}  \caption[]{\small
  Graphical representation of the formula for the contributions missing from the linear term in the recursion relation. The variable $b_1$ is understood to be summed over the range $4\leq b_1 \leq n-1$.}
  \label{missingpathstree}
\end{figure}

So let us examine the quadratic term in (\ref{fullPrec}). We begin by looking at the summand.
The first two factors $R_{n;2\,i}\mathcal{P}_i(2,\ldots,-\hat{P},\hat{1})$ taken together reproduce a sum over vertical paths in a tree very similar to the one in Fig. \ref{fig-rec-solution}. The relevant tree diagram is shown in the left half of Fig. \ref{fig-rec-product} (ignoring the solid lines for now).
Let us describe the differences between this tree and the one of Fig. \ref{fig-rec-solution}. Firstly, the root vertex corresponds to $R_{n;2\,i}$ instead of 1, so the first term in the sum over paths, 1, is absent. Secondly, the top $R$-invariant $R_{n;2\,i}$ has its labels fixed to be 2 and $i$. Thirdly, all descendant vertices have at least two pairs of labels due to the fact that the last leg of the argument of $\mathcal{P}_i(2,\ldots,-\hat{P},\hat{1})$ is $\hat{1}$ and not $n$. As we saw in equations (\ref{remove1hat}) and (\ref{extension}) this results in the replacement $\l< n| \rightarrow \l< n|x_{ni}x_{i2}$ which induces extra labels on the $R$-invariants. Finally, the right-most vertex of each descendant cluster has two pairs of indices. Thus the right superscripts associated to these vertices are not 0, as was the case for the tree in Fig. \ref{fig-rec-solution}. Instead these superscripts are all $2\,i$ which is obtained by deleting the final pair and reversing the order of the remaining pair $i\,2$.

\begin{figure}
\psfrag{one}[cc][cc]{$1$}
\psfrag{a1b1}[cc][cc]{$2\,i$}
\psfrag{a2b2}[cc][cc]{$c_{1}d_{1}$}
\psfrag{a3b3}[cc][cc]{$c_{2}d_{2}$}
\psfrag{b1a1a2b2}[cc][cc]{$i\,2;a_{2}b_{2}$}
\psfrag{b2a2a3b3}[cc][cc]{$d_{1}c_{1};c_{2}d_{2}$}
\psfrag{b1a1a3b3}[cc][cc]{$i\,2;a_{3}b_{3}$}
\psfrag{b1a1b2a2a3b3}[cc][cc]{$i\,2;b_{2}a_{2};a_{3}b_{3}$}
\psfrag{two}[cc][cc]{$2$}
\psfrag{n1}[cc][cc]{\,\,\,\,$n-1$}
\psfrag{a1p}[cc][cc]{$3$}
\psfrag{a2p}[cc][cc]{\!\!$a_{2}+1$}
\psfrag{c2p}[cc][cc]{\!\!\!\!\!\!$c_{1}+1$}
\psfrag{b1}[cc][cc]{$i$}
\psfrag{b2}[cc][cc]{$b_{2}$}
\psfrag{d2}[cc][cc]{$d_{1}$}
\psfrag{times}[cc][cc]{$\times$}
 \centerline{{\epsfysize5cm
\epsfbox{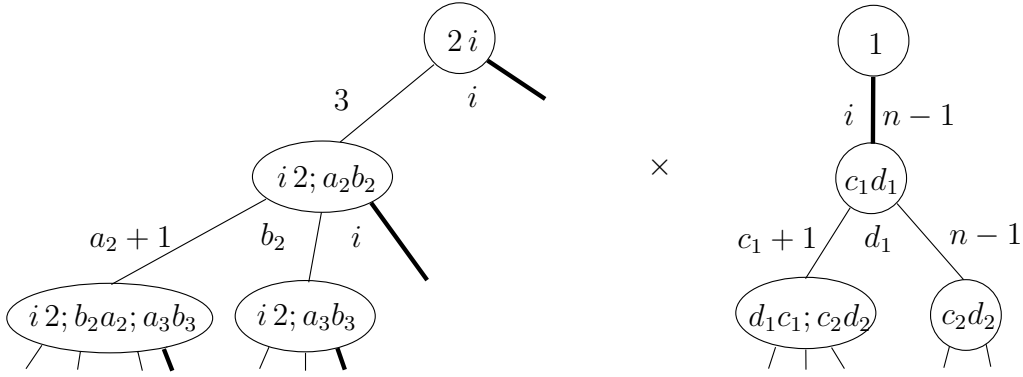}}}  \caption[]{\small
  Graphical representation of the quadratic term $R_{n;2\,i} \mathcal{P}_i(2,\ldots,-\hat{P_i},\hat{1}) \mathcal{P}_{n-i+2}(\hat{P_i},i,\ldots,\bar{n})$ in equation (\ref{fullPrec}).
The left tree corresponds to the first two factors $R_{n;2\,i} \mathcal{P}_i(2,\ldots,-\hat{P_i},\hat{1}) $, while the right tree corresponds to the final factor $\mathcal{P}_{n-i+2}(\hat{P_i},i,\ldots,\bar{n})$. As indicated in the text, after summing over $i$ the first tree is almost what is needed to complete the linear term to $\mathcal{P}_n$. The missing pieces come from the right factor which can be adjoined to the left by inserting it everywhere there is a line drawn in bold so that these lines then all lead to a descendant vertex with labels $c_2,d_2$. Since the $c$ and $d$ labels are all dummy variables they can then be exchanged for the suitable $a$ and $b$ labels by a change of notation.}
  \label{fig-rec-product}
\end{figure}

Now let us consider the sum over vertical paths in the tree diagram we have just described. There is one path of length 0, corresponding to the contribution,
\be
R_{n;2i}\,.
\label{left0}
\ee
There is one path of length one which gives
\be
R_{n;2i} \sum_{3\leq a_2,b_2 \leq i} \!\!\!\!\!\! R_{n;i2;a_2b_2}^{0;2i} \, .
\label{left1}
\ee
There are two paths of length two which give the following two contributions,
\be
R_{n;2i} \sum_{3\leq a_2,b_2 \leq i} \!\!\!\!\!\! R_{n;i2;a_2b_2}^{0;2i} \Bigl[\sum_{a_2+1 \leq a_3,b_3} \!\!\!\!\!\! R_{n;i2;b_2a_2;a_3b_3}^{0;i2a_2b_2} + \sum_{b_2 \leq a_3,b_3 \leq i} \!\!\!\!\!\! R_{n;i2;a_3b_3}^{i2a_2b_2;2i} \Bigr] \, .
\label{left2}
\ee
Continuing, we have five paths of length three and so on. Since we consider the sum over paths, we have to add up all these terms.

Now we consider the third factor $\mathcal{P}_{n-i+2}(\hat{P_i},i,\ldots,\bar{n})$ in the summand of the quadratic term in (\ref{fullPrec}). This gives us the sum over paths in the tree shown in the right half of Fig. \ref{fig-rec-product}. This tree is again similar to the tree shown in Fig. \ref{fig-rec-solution}. There are two differences between this tree and the one of Fig. \ref{fig-rec-solution}. Firstly, the outermost lower summation limit is $i$ and not 2. Secondly, since the first argument of $\mathcal{P}_{n-i+2}(\hat{P_i},i,\ldots,\bar{n})$ is $\hat{P}_i$ and not $i-1$, there will be a non-trivial left superscript associated to the first descendant vertex. This is precisely the same effect that we saw in equations (\ref{removePhat}) and (\ref{Phatrep}). The corresponding superscript is $2\,i$ so that the first descendant vertex corresponds to $R_{n;c_1d_1}^{2i;0}$.

Writing out the terms in the sum over paths in the tree corresponding to $\mathcal{P}_{n-i+2}(\hat{P},i,\ldots,\bar n)$ we find from paths of length 0,
\be
1\, ,
\label{right0}
\ee
from paths of length one,
\be
\sum_{i\leq c_1,d_1 \leq n-1} \!\!\!\!\!\! R_{n;c_1d_1}^{2i;0}\, ,
\label{right1}
\ee
from paths of length two,
\be
\sum_{i\leq c_1,d_1 \leq n-1} \!\!\!\!\!\! R_{n;c_1d_1}^{2i;0} \Bigl[ \sum_{c_1+1 \leq c_2,d_2 \leq d_1} \!\!\!\!\!\! R_{n;d_1c_1;c_2d_2}^{0;c_1d_1} + \sum_{d_1 \leq c_2,d_2 \leq n-1} \!\!\!\!\!\! R_{n;c_2d_2}^{c_1d_1;0} \Bigr],
\label{right2}
\ee
and so on.

Thus the left half of Fig. \ref{fig-rec-product} gave us the sum of (\ref{left0}), (\ref{left1}), (\ref{left2}) and longer paths. The right half of Fig. \ref{fig-rec-solution} gave us the sum of (\ref{right0}), (\ref{right1}), (\ref{right2}) and longer paths. If we consider the product of the expressions obtained from the two trees we see that it can be written,
\begin{align}
&R_{n;2i} \notag \\
+&R_{n;2i} \Bigl[\sum_{3\leq a_2,b_2 \leq i} \!\!\!\!\!\! R_{n;i2;a_2b_2}^{0;2i} + \sum_{i\leq c_1,d_1 \leq n-1} \!\!\!\!\!\! R_{n;c_1d_1}^{2i;0}\Bigr] \notag \\
+&R_{n;2i} \Biggl[ \sum_{3\leq a_2,b_2 \leq i} \!\!\!\!\!\! R_{n;i2;a_2b_2}^{0;2i} \Bigl[\sum_{a_2+1 \leq a_3,b_3} \!\!\!\!\!\! R_{n;i2;b_2a_2;a_3b_3}^{0;i2a_2b_2} + \sum_{b_2 \leq a_3,b_3 \leq i} \!\!\!\!\!\! R_{n;i2;a_3b_3}^{i2a_2b_2;2i} + \sum_{i\leq c_1,d_1 \leq n-1} \!\!\!\!\!\! R_{n;c_1d_1}^{2i;0} \Bigr] \notag\\
&\phantom{R_{n;2i}} +  \sum_{i\leq c_1,d_1 \leq n-1} \!\!\!\!\!\! R_{n;c_1d_1}^{2i;0} \Bigl[ \sum_{c_1+1 \leq c_2,d_2 \leq d_1} \!\!\!\!\!\! R_{n;d_1c_1;c_2d_2}^{0;c_1d_1} + \sum_{d_1 \leq c_2,d_2 \leq n-1} \!\!\!\!\!\! R_{n;c_2d_2}^{c_1d_1;0} \Bigr]\Biggl] + \text{ longer .}
\end{align}
Remembering that we need to sum over $i$ in the quadratic term on the RHS of (\ref{fullPrec}), we find precisely the terms we are looking for.
To make the identification completely explicit we can perform the changes of labels $c_1,d_1 \rightarrow a_2,b_2$ in the second line, $c_1,d_1 \rightarrow a_3,b_3$ in the third line and $c_1,d_1 \rightarrow a_2,b_2, \,\, c_2,d_2 \rightarrow a_3,b_3$ in the fourth line, and finally rename the summation variable $i$ as $b_1$. This analysis can also be seen diagrammatically. If one glues the tree from the right half of Fig. \ref{fig-rec-product} to that from the left half everywhere there is a line drawn in bold and performs the corresponding changes of labels, one obtains exactly the tree diagram of Fig. \ref{missingpathstree}. 

Thus finally we arrive at the fact that the sum of the linear term and quadratic term on the RHS reproduces the sum over vertical paths in Fig. \ref{fig-rec-solution}. This completes the inductive step of the proof. It remains to note that the sum over paths in Fig. \ref{fig-rec-solution} coincides with the first few amplitudes as we have seen by considering NMHV and NNMHV amplitudes. Therefore we conclude that formula (\ref{paths}) does indeed produce the full tree-level super-amplitude.


\section{Symmetries of the amplitudes}\label{sect-app-symmetry}

Tree amplitudes in $\mathcal{N}=4$ SYM are expected to have many
symmetries. First of all, $\mathcal{N}=4$ SYM is a superconformal
field theory, so the amplitudes should exhibit this symmetry in their
functional forms. The MHV super-amplitudes were shown to be
annihilated by all generators of the conventional superconformal
algebra in \cite{Witten:2003nn}. The amplitudes we have constructed in
this paper are manifestly invariant under all generators of the
conventional superconformal algebra\footnote{Following the conventions
  of \cite{dhks5} we will use lower case characters to denote the
  conventional superconformal generators and upper case ones for the
  dual superconformal generators.} except for the superconformal
symmetries $s,\bar s, k$.

 In addition to the conventional superconformal symmetry, it was
 conjectured in \cite{dhks5} that the tree-level super-amplitudes
 should also exhibit {\it dual} superconformal symmetry. As far as
 tree-level super-amplitudes are concerned, the conjecture of
 \cite{dhks5} states that they should be covariant under dual
 conformal transformations $K$ and the chiral superconformal
 transformations $S$, while they are invariant under $P,Q,\bar Q,\bar
 S$. They also have the obvious property that the dual dilatation
 weight and central charge are equal to $n$, the number of particles.

The generators of the two different realisations of the superconformal
algebra are not all independent. As discussed in \cite{dhks5} the odd
generator $\bar q$ coincides with $\bar S$, while $\bar s$ coincides
with $\bar Q$. The same correspondence was observed in
\cite{Berkovits:2008ic,Beisert:2008iq} after performing a fermionic
T-duality in the string sigma model. The explicit form of all
generators is summarised in Appendix \ref{app-generators}.

In \cite{Brandhuber:2008pf} the dual conformal covariance of the
tree-level super-amplitudes was verified recursively using the
supersymmetric recursion relations. We can indeed see this symmetry in
the explicit form of the solution we have presented. All quantities
$R_{n;a_1b_1;...;a_mb_m;st}$ are dual conformal invariants, as can
be quickly verified by counting the conformal weights of the numerator
and denominator. For tree-level amplitudes, this is sufficient to show
dual superconformal covariance, as claimed in
\cite{Brandhuber:2008pf}, since the conventional
superconformal invariance $\bar s \mathcal{A}=0$ of the amplitude
should be unbroken. In
other words if we know that $\bar s \mathcal{A} = 0$ then we have $\bar
Q \mathcal{A} =0$, and together with covariance under dual inversions
this is sufficient to derive all the expected
properties under the full dual superconformal algebra. Further we
remark that if all super-amplitudes obey $\bar s \mathcal{A} =0$ then
they also obey $s \mathcal{A} =0$, since we could alternatively have
performed the entire analysis in the anti-chiral ($\bar \eta$)
representation for the gluon supermultiplet. Thus showing $\bar
s$-invariance is sufficient to derive invariance under $s$ and
therefore under $k=\{s,\bar s\}$.

In general, showing the conventional superconformal invariance of the
tree-level amplitudes is a non-trivial task (see
e.g. \cite{Witten:2003nn}). Here we will explicitly show that
expression (\ref{paths}) does indeed obey this symmetry.
As we have seen, the only property of the super-amplitude which
remains to be explicitly verified is its behaviour
under the $\bar s^A_{\dot\alpha}$ or $\bar{Q}^{A}_{\dot{\alpha}}$
supersymmetry. We recall the explicit form of $\bar{Q}^{A}_{\dot{\alpha}}$,
\begin{equation}\label{recallQbar}
\overline{Q}_{\dot{\alpha}}^A = \sum_i [\theta_i^{\alpha A}
  \partial_{i \alpha \dot{\alpha}} + \eta_i^A \partial_{i \dot{\alpha}}]\,.
\end{equation}
The invariance of the NMHV super-amplitude (\ref{PNMHV}) was shown in \cite{dhks5}. It follows from the fact that\footnote{We omit the factor $\langle 12 \rangle \ldots \langle n 1 \rangle$ in the denominator of the superamplitude since this is obviously invariant under the action of $\overline{Q}$.}
\begin{equation}\label{QbarNMHV}
\bar{Q}^{A}_{\dot{\alpha}}\, \delta^{(4)}(p)\, \delta^{(8)}(q) \, R_{n;a_{1}b_{1}} = 0\,.
\end{equation}
Following \cite{dhks5}, we can simplify calculations such as (\ref{QbarNMHV})
by noting that the super-amplitudes are invariant under $\bar{S}_{\dot{\alpha} A}$ and $Q_{\alpha A}$.
Since translations, Lorentz rotations and the combination $D-C$ are also symmetries of the super-amplitudes, we have (see Appendix \ref{app-generators})
$\left\{ \bar{Q}^{\dot\alpha A}, \bar{S}_{\dot\beta B} \right\} \mathcal{A}_{n} = \left\{ \bar{Q}^{\dot\alpha A}, {Q}_{\alpha B} \right\} \mathcal{A}_{n} = 0$.
This allows us to compute the variation $\bar{Q}^{A}_{\dot{\alpha}} \mathcal{A}_{n}$ in a frame obtained by a combined shift using $\bar{S}_{\dot{\alpha} A}$ and $Q_{\alpha A}$.
In particular, we can choose the shift parameters such that $\theta_{a_{1}}=\theta_{b_{1}}=0$ \cite{dhks5}.

Let us proceed with the NNMHV super-amplitude (\ref{PNNMHVnew}).
We first consider terms in (\ref{PNNMHVnew}) which are not affected by boundary effects, i.e. 
$R_{n;a_{1}b_{1}} R_{n;a_{2}b_{2}}$ and $R_{n;a_{1}b_{1}} R_{n;b_{1}a_{1};a_{2}b_{2}}$.
From (\ref{QbarNMHV}) we immediately see that the terms with $R_{n;a_{1}b_{1}} R_{n;a_{2}b_{2}}$ are invariant under $\bar{Q}^{A}_{\dot{\alpha}}$.
Let us now consider the variation
\begin{equation}\label{QbarNNMHV1}
 \bar{Q}^{A}_{\dot{\alpha}}\, \delta^{(4)}(p) \,\delta^{(8)}(q)\, R_{n;a_{1}b_{1}} R_{n;b_{1}a_{1};a_{2}b_{2}} =  \delta^{(4)}(p)\, \delta^{(8)}(q)\, R_{n;a_{1}b_{1}} \,\bar{Q}^{A}_{\dot{\alpha}}\,  R_{n;b_{1}a_{1};a_{2}b_{2}}\,.
\end{equation}
Following \cite{dhks5}, we can choose a fixed frame in which
$\theta_{a_{2}}=\theta_{b_{2}}=0$.
In this frame, (\ref{generalR}) simplifies to 
\be \label{Xi-fixedframe}
\Xi^{A}_{n;b_{1}a_{1};a_{2}b_{2}} = x_{a_{2}b_{2}}^2 \langle \xi 
\theta^{A}_{a_{1}} \rangle  
\ee
and
\be 
\label{app-newR1} 
R_{n;b_{1}a_{1};a_{2}b_{2}} =  \frac{1}{4!}\epsilon_{ABCD}
\frac{\vev{\xi\,\theta_{a_{1}}^{A}}\vev{\xi\,\theta_{a_{1}}^{B}}\vev{\xi\,\theta_{a_{1}}^{C}}
\vev{\xi\,\theta_{a_{1}}^{D}}}{\vev{\xi
\, I_{1}} \vev{\xi \, I_{2}}\vev{\xi \, I_{3}}\vev{\xi \,
I_{4}}}\, (x_{a_{2}b_{2}}^2)^3 \langle a_{2}\, a_{2}-1 \rangle \langle b_{2}\, b_{2}-1 \rangle\,.
\ee 
Here 
\be
|I_{1}\rangle = x_{a_{1}a_{2}} x_{a_{2}b_{2}} | b_{2} \rangle,\,
|I_{2}\rangle = x_{a_{1}a_{2}} x_{a_{2}b_{2}} | b_{2}-1 \rangle,\, |I_{3}\rangle = x_{a_{1}b_{2}} x_{b_{2}a_{2}} | a_{2} \rangle,\, |I_{4}\rangle = x_{a_{1}b_{2}} x_{b_{2}a_{2}} | a_{2}-1 \rangle
\ee
and
\be
\langle \xi | = \langle n | x_{n b_1} x_{b_1 a_1}.
\ee
Further, when computing the  $\bar{Q}^{A
\, \dot{\alpha}}$-variation of $R_{n;b_{1}a_{1};a_{2}b_{2}}$ in (\ref{app-newR1})
we can drop all terms in (\ref{recallQbar}) except $\theta_{n\, \alpha}^{A}
\partial^{\alpha \dot{\alpha}}_{n}+\theta_{a_{1}\, \alpha}^{A}
\partial^{\alpha \dot{\alpha}}_{a_{1}}+\theta_{b_{1}\, \alpha}^{A}
\partial^{\alpha \dot{\alpha}}_{b_{1}}$. The reason is that there is no explicit dependence on $\tilde\lambda$ in  $R_{n;b_{1}a_{1};a_{2}b_{2}}$, and
that $\theta_{a_{2}\, \alpha}^{A}
\partial^{\alpha \dot{\alpha}}_{a_{2}}=\theta_{b_{2}\, \alpha}^{A}
\partial^{\alpha \dot{\alpha}}_{b_{2}}=0$ in the fixed frame.
Let $\tilde{\lambda}_{J}^{\dot{\alpha}}$ be an arbitrary projection.
It can be easily seen that in the fixed frame, $\lbrack J
\bar{Q}^{E} \rbrack$ acts trivially on $I_{i}$ in
(\ref{app-newR1}), because e.g. \be \langle \xi \lbrack J
\bar{Q}^{E} \rbrack I_{1} \rangle = \langle \xi \, \theta_{a_{1}}^{E}
\rangle \lbrack J | x_{a_{2}b_{2}} | b_{2} \rangle \ee is annihilated by the
Grassmann delta function in the numerator of (\ref{app-newR1}).
Thus when acting with $\lbrack J \bar{Q}^{E} \rbrack$ on
(\ref{app-newR1}), only $\langle \xi |$ transforms. After using
the cyclic identity for spinors we easily obtain
\be \label{app-newR2}\lbrack J \bar{Q}^{E} \rbrack
\,R_{n;b_{1}a_{1};a_{2}b_{2}} = \frac{1}{4!} \epsilon_{ABCD}  \frac{\chi^{A}
\vev{\xi\,\theta_{a_{1}}^{B}}\vev{\xi\,\theta_{a_{1}}^{C}}\vev{\xi\,\theta_{a_{1}}^{D}}
}{\vev{\xi \, I_{1}} \vev{\xi \, I_{2}}\vev{\xi \, I_{3}}\vev{\xi
\, I_{4}}} \times \left\lbrack
\l<n|x_{nb_{1}}x_{b_{1}a_{1}}|\theta^{E}_{a_{1} n}\r> +
\l<n|x_{n a_{1}}x_{a_{1}b_{1}}|\theta^{E}_{b_{1}n}\r> \right\rbrack \,,\ee
where the explicit expression for $\chi^{A}$ is inessential to our argument.
From (\ref{app-newR2}) we see that $R_{n;b_{1}a_{1};a_{2}b_{2}}$ is {\it not} dual superconformally
invariant. However, in (\ref{PNNMHVnew}), it always appears multiplied
by the invariant $R_{n;a_{1}b_{1}}$. In this case, the Grassmann
delta function in $R_{n;a_{1}b_{1}}$ makes the variation
(\ref{app-newR2}) vanish, and therefore $R_{n;a_{1}b_{1}}
R_{n;b_{1}a_{1};a_{2}b_{2}}$ is a dual superconformal invariant. The boundary terms in the sums behave in a similar way. The replacement spinors produce additional terms in the $\bar Q$ variation which are annihilated by the presence of the Grassmann factors.

We conclude that the NNMHV amplitudes are dual superconformally
covariant. From the discussion here and in section \ref{sect-all}
it is easy to see that this property is true for all tree-level
amplitudes in $\mathcal{N}=4$ SYM. Indeed, one can repeat the
argument above to `longer' chains of invariants that appear in equation (\ref{paths}).
Take for example $R_{n;a_{1}b_{1}} R_{n;b_{1}a_{1};a_{2}b_{2}} R_{n;b_{1}a_{1};b_{2}a_{2};a_{3}b_{3}}$ from (\ref{NNNMHV-1}).
After fixing a frame where $\theta_{a_{3}}=\theta_{b_{3}}=0$, we obtain an expression like
(\ref{app-newR1}) with a different $\langle \xi | = \langle n | x_{n b_{1}} x_{b_{1} a_{1}} x_{a_{1} b_{2}} x_{b_{2}a_{2}}|$.
Because of the linearity of $\lbrack J \bar{Q}^{E} \rbrack$ the calculation of the variation
of $R_{n;b_{1}a_{1};b_{2}a_{2};a_{3}b_{3}}$ is as above, except that now we obtain two contributions, one of which vanishes
thanks to $R_{n;a_{1}b_{1}}$, and the other thanks to $R_{n;b_{1}a_{1};a_{2}b_{2}}$. The crucial feature is that $R$'s with
many indices share all first indices of their `predecessors'. This is the case by construction for
all terms in (\ref{paths}).

Therefore we have shown explicitly that the formula (\ref{paths}) for all
tree-level amplitudes in $\mathcal{N}=4$ SYM has all the expected
properties under both conventional and dual superconformal symmetry.

\section{Gluon scattering amplitudes from super-amplitudes}\label{sect-app-components}
Here we wish to give some explanations on how gluon amplitudes can be extracted from
our solutions and how this can be implemented, for example on a computer.

Let us first stress that any component amplitudes for arbitrary particle or helicity choice
can be extracted from the super-amplitudes, see e.g. \cite{dhks5} for more explanations. Here we focus on the
particularly simple case of gluon amplitudes.

According to (\ref{super-wave}), to each negative helicity gluon at position $j$ is associated a factor of $(\eta_{j})^4 = \eta_{j}^{1} \eta_{j}^{2} \eta_{j}^{3} \eta_{j}^{4}$, and to each positive helicity gluon simply a factor of $1$. Going from a given super-amplitude to a gluon component amplitude therefore just amounts to extracting
specific prefactors in the $\eta$-expansion of the super-amplitude.
An elementary example is the relation (\ref{intro-expansion}) between the gluon MHV amplitude (\ref{intro-MHV-gluons})
and the super-amplitude (\ref{intro-MHV-n}).

A less trivial example is the split-helicity NMHV amplitude,
\be \label{app-expansion} {\cal A}^{\rm NMHV}_{n} =
\left(\eta_{n-2}\right)^4 \left(\eta_{n-1}\right)^4 \left(\eta_{n}\right)^4
A(1^{+},\ldots,(n-3)^{+},(n-2)^{-},(n-1)^{-},n^{-}) + \ldots\,, \ee
We want to  expand ${\cal A}^{\rm NMHV}_{n}$ in $\eta$ and recover the desired split-helicity gluon amplitude.\footnote{Note that a Grassmann delta function is simply defined as a product,
$\delta^{(4)}(\chi^{A}) = {1}/{4!} \epsilon_{ABCD} \chi^{A} \chi^{B} \chi^{C} \chi^{D}$.}
A simple way to achieve this is to observe that the relation between
NMHV super-amplitude and the desired gluon component can be written as
a Grassmann integral
\be \label{gluons-etaintegral}
A(1^{+},\ldots,(n-3)^{+},(n-2)^{-},(n-1)^{-},n^{-}) = \int d^{4}\eta_{n-2} \int d^{4}\eta_{n-1} \int d^{4}\eta_{n} \,{\cal A}^{\rm NMHV}_{n}\,.
\ee
In this paper we have already encountered many such Grassmann integrals and seen that they are easy to do.
We can always choose two arbitrary spinor projections of $q^{A}_{\alpha}$ to rewrite the $\delta^{(8)}(q^{A}_{\alpha})$ of ${\cal A}^{\rm NMHV}_{n}$ as
\be \label{app-delta8}
\delta^{(8)}(q^{A}_{\alpha}) =
 \vev{n-1 \,n}^4 \,
\delta^{(4)}\left( \eta_{n-1}^{A} + \sum_{i=1}^{n-2} \frac{\vev{i n}}{\vev{n-1\,n}} \eta_{i}^{A} \right) \,
\delta^{(4)} \left( \eta_{n}^{A} + \sum_{i=1}^{n-2} \frac{\vev{n-1 \,i}}{\vev{n-1\,n}} \eta_{i}^{A} \right)\,.
\ee
This allows us to immediately carry out the $d^{4}\eta_{n}$ and $d^{4}\eta_{n-1}$ integrals in (\ref{gluons-etaintegral}). The remaining terms in ${\cal A}^{\rm NMHV}_{n}$ are unaffected by this since they
can be written in the form (\ref{Xi-eta2}) in which they are independent of $\eta_{n-1}$ and $\eta_{n}$.
Hence we obtain
\be
A(1^{+},\ldots,(n-3)^{+},(n-2)^{-},(n-1)^{-},n^{-}) = \delta^{(4)}(p)\,\frac{ \vev{n-1 \,n}^4}{\vev{12}\ldots \vev{n1}} \int d^{4}\eta_{n-2} \sum_{1<s,t<n} R_{n;s,t}\,,
\ee
where the $\Xi_{n;s,t}$ in $R_{n;s,t}$ are written in the form (\ref{Xi-eta2}).
A further simplification occurs because $R_{n;s,t}$ only depends on $\eta_{n-2}$ if $t=n-1$, see (\ref{Xi-eta2}). Carrying out the remaining Grassmann integration using the $\delta^{(4)}(\Xi_{n;s,n-1})$ in $R_{n;s,n-1}$ we obtain
\ba
A(1^{+},\ldots,(n-3)^{+},(n-2)^{-},(n-1)^{-},n^{-}) &=&   \nonumber \\
&& \hspace{-7 cm} -\frac{\delta^{(4)}(p)}{\vev{12}\ldots
\vev{n-3\,n-2} \l< n1\r>}  \sum_{s=2}^{n-3}\frac{\langle
n-2|x_{n-1\,s}x_{sn}|n\rangle^3 \vev{s\,s-1}} {x_{s\,n-1}^2
x_{s\,n}^2 \lbrack n-1 | x_{n-1,s} |s\rangle\, \lbrack n-1 |
x_{n-1\,s}|s-1\rangle}\,. \ea This is in perfect agreement with
formula (4.5) given in \cite{Britto:2005dg}.

We can continue further and derive, for example, a formula for the split-helicity NNMHV amplitudes. Just as in the NMHV case, we can write all invariants so that they do not depend on $\eta_{n}$ or $\eta_{n-1}$. Then performing integrals with respect to these variables just produces a factor of $\l< n-1\, n\r>^4$ from the $\delta^8(q)$ factor. The remaining integrals with respect to $\eta_{n-2}$ and $\eta_{n-3}$ give nothing from the second term in (\ref{PNNMHVnew}). From the first term in (\ref{PNNMHVnew}) we obtain two contributions, one where $b_1=n-1$ and $b_2=n-2$ and one where $b_1=b_2=n-1$. The final formula for the gluon amplitudes is
\begin{align}
A(1^+,\ldots,(n-4)^+,(n-3)^-,(n-2)^-,(n-1)^-,n^-) = \delta^{(4)}(p) (S_1 + S_2),
\end{align}
where the two terms are given by
\begin{align}
S_1 &= \frac{\l<n\, n-1\r> \l<n-1 \, n-2\r> \l< n-2 \, n-3\r>}{\prod_{i=1}^n \l< i \, i+1 \r>}\sum_{a_1=2}^{n-5} \sum_{a_2=a_1+1}^{n-4} \frac{N_1}{D_1}, \\
S_2 &= \frac{\l<n \, n-1\r> \l<n-1 \, n-2\r> \l<n-2 \, n-3\r>^4}{\prod_{i=1}^n \l< i \, i+1 \r>}\sum_{a_1=2}^{n-4} \sum_{a_2=a_1+1}^{n-3} \frac{N_2}{D_2}.
\end{align}
Here the numerators and denominators of the summands are
\begin{align}
N_1 = &\l<a_1 \, a_1-1\r>  \l<n |x_{na_1} x_{a_1 n-1}|n-2\r>^3 \l< a_2 \,a_2-1\r>  [n-1|x_{n-1 a_1}x_{a_1 a_2}x_{a_2 n-3} |n-3\r>^3,
\\ D_1=&[n-1|x_{n-1 a_1}|a_1\r> [n-1| x_{n-1 a_1}|a_1-1\r>  [n-1|x_{n-1 a_1}x_{a_1a_2}x_{a_2n-2}|n-2\r> \notag\\
&[n-1|x_{n-1 a_1} x_{a_1 n-2} x_{n-2 a_2}|a_2\r> [n-1|x_{n-1 a_1} x_{a_1 n-2} x_{n-2 a_2}|a_2-1\r>x_{a_1 n-1}^2 x_{n a_1}^2 x_{a_2 n-2}^2,\\
\notag \\
N_2 =&\l <a_1 a_1-1\r> \l<a_2 a_2-1\r> \l<n|x_{na_1} x_{a_1 n-1} x_{n-1 a_2} x_{a_2 a_1} x_{a_1 n-1} | n-1]^3 ,\\
D_2 =&[n-1|x_{n-1 a_1}|a_1\r> [n-1|x_{n-1 a_1}|a_1-1\r>  [n-1|x_{n-1 a_1} x_{a_1 a_2} x_{a_2 n-1} |n-2 \r>\notag\\
&[n-1|x_{n-1 a_2} |a_2\r> [n-1|x_{n-1 a_2}|a_2-1\r> (x_{a_1 n-1}^2)^3 x_{a_2 n-1}^2 x_{n a_1}^2.
\end{align}

It is simple to check analytically that this formula correctly reproduces the six-point $\bMHV$ amplitude and the seven-point next-to-$\bMHV$ amplitude. We have also checked numerically that it coincides with the six terms given in \cite{Britto:2005dg} for the eight-point NNMHV split-helicity gluon amplitude.

In more complicated situations one could for example first do some
$\eta$ integrations analytically
(e.g. using the $\delta^{(8)}(q)$ which is present in all physical
super-amplitudes because of supersymmetry),
and then implement the remaining integrations/expansions on a computer.
This can be easily programmed, keeping track of the overall sign
(because the $\eta$'s are anticommuting variables).
The resulting spinor expressions can be evaluated numerically using
available packages, see e.g. \cite{Maitre:2007jq}.

\section{Conclusions}\label{sect-conclusions}

The main result of our paper is formula (\ref{paths}) for all
tree-level amplitudes in $\mathcal{N}=4$ SYM. The formula contains all
amplitudes with arbitrary total helicity (MHV,NMHV,...,$\bMHV$).
It is given in terms of vertical paths of a particular rooted tree,
shown in Fig. \ref{fig-rec-solution}.
This extends previous solutions of the BCF recursion relations which
applied only to the closed subset of split-helicity gluon amplitudes
\cite{Britto:2005dg}.
Our solution is written in on-shell $\mathcal{N}=4$ superspace. It is
built from dual superconformal invariants and so it manifestly
exhibits both conventional and dual superconformal symmetries.

Our expression contains as components all amplitudes for arbitrary external
states and helicities. We explained in section
\ref{sect-app-components} that gluon components are particularly
simple to extract, since they can be obtained from the
super-amplitudes by carrying out
Grassmann integrations. A crucial simplifying feature is that
(\ref{paths}) is built from
sums over products of Grassmann delta functions, which can be used to
perform the aforementioned
integrations. We expect that it will be possible to obtain compact
expressions for previously
unknown gluon components following the example in section
\ref{sect-app-components}.

We expect our results to be relevant for $\mathcal{N}=8$ supergravity
as well, since tree-level amplitudes
in the latter theory can be obtained from those in $\mathcal{N}=4$ SYM
through the KLT relations \cite{Kawai:1985xq}. Furthermore the methods
employed here could also be directly applied to solving recursion
relations for supergravity tree-level amplitudes
\cite{ArkaniHamed:2008yf}. It would also be interesting to see if our
formula could shed light on the relation among tree-level amplitudes
described in  \cite{Bern:2008qj}.

\section*{Acknowledgements}

We are grateful to Nathan Berkovits, Paul Heslop, Juan Maldacena, Radu
Roiban and particularly Gregory Korchemsky and Emery Sokatchev for
interesting and stimulating discussions. This research was supported in part by the French Agence Nationale de la
Recherche under grant ANR-06-BLAN-0142.

\section*{Appendices}

\appendix

\section{Collinear limit of the super-amplitudes}\label{app-collinear}

Here we check that our amplitudes have the correct collinear limit
as two particles become almost collinear \cite{Mangano:1990by}.
Consider two neighbouring particles at points $a$ and $b=a+1$ that
become collinear such that \be p_{a} = z P \,,\qquad p_{b} = (1-z)
P\,, \ee then an $n$-gluon tree amplitude is expected to behave as
\be \label{coll-components} A_{n} \stackrel{a
||b}{\longrightarrow} \sum_{\lambda=\pm} {\rm Split}^{\rm
tree}_{-\lambda}(a^{\lambda_{a}},b^{\lambda_{b}}) \,A_{n}(\ldots
\,,(a+b)^{\lambda}\,, \ldots)\,, \ee where ${\rm Split}^{\rm
tree}_{-\lambda}$ are certain helicity-dependent splitting
functions, see \cite{Mangano:1990by}. The non-vanishing splitting
functions diverge as $1/\sqrt{s_{ab}}$ in the collinear limit
$s_{ab}=(p_{a}+p_{b})^2 \to 0$. In the collinear limit, the
spinors corresponding to the momenta $p_{a}$ and $p_{b}$ become
\be\label{coll-spinors} \lambda_{a} \to \sqrt{z}
\lambda_{P}\,,\qquad \tilde{\lambda_{a}} \to \sqrt{z}
\tilde{\lambda}_{P}\,,\qquad \lambda_{b} \to \sqrt{1-z}
\lambda_{P}\,,\qquad \tilde{\lambda_{b}} \to \sqrt{1-z}
\tilde{\lambda}_{P}\,. \ee In the supersymmetric case, to be
consistent with (\ref{coll-spinors}) we also define\be \eta_{a} \to \sqrt{z}
\eta_{P}\,,\qquad \eta_{b} \to \sqrt{1-z} \eta_{P}\,. \ee

By inspecting the collinear limit for the MHV super-amplitudes
(\ref{intro-MHV-n}), we expect the following collinear limit for
super-amplitudes at tree level, \be \label{coll-super}
\mathcal{A}_{n}(\ldots, a,b,\ldots) \stackrel{a
||b}{\longrightarrow} \frac{1}{\sqrt{z (1-z)} \vev{ab}}
\mathcal{A}_{n-1}(\ldots \,,P\,, \ldots)\,. \ee

Let us see if relation (\ref{coll-super}) holds for the NMHV
amplitudes (\ref{intro-NMHV}) as well. We need to analyse the
behaviour of the invariants $R_{n;s,t}$ in the limit. Because of
cyclic symmetry of the super-amplitude, we can consider the
$a=n-1,b=n$ without loss of generality. This is advantageous
because then the invariants $R_{n;s,t}$ are affected by the
collinear limit only through  $\lambda_{n} = \sqrt{1-z}
\lambda_{P}$. Looking at (\ref{def-Rrst}) we see that \be
\label{coll-NMHV1} R_{n;s,t} \stackrel{n-1 ||n}{\longrightarrow}
R_{P;s,t}\,. \ee We also observe that \be\label{coll-NMHV2}
R_{n;s,n-1}  \stackrel{n-1 ||n}{\longrightarrow} R_{P;s,n-1}
\propto \vev{n-1\,n}^2 \to 0\,. \ee Using (\ref{coll-NMHV1}) and
(\ref{coll-NMHV2}) on (\ref{intro-NMHV}) we see that indeed \be
\mathcal{A}^{\rm NMHV}_{n}(1,\ldots,n-1,n)
 \stackrel{n-1 ||n}{\longrightarrow} \frac{1}{\sqrt{z (1-z)}
\vev{n-1\,n}} \mathcal{A}^{\rm NMHV}_{n-1}(1,\ldots,n-2,P)\,. \ee
Going to NNMHV amplitudes (\ref{ANNMHV},\ref{PNNMHVnew}), we see that the behaviour of the `longer' invariants like
$R_{n;u,v}R_{n;v,u;s,t}$
under the collinear limit where particles $n-1$ and $n$ become collinear
is completely analogous to the NMHV case, they turn into $R_{P;u,v}R_{P;v,u;s,t}$. It is then
obvious that (\ref{ANNMHV},\ref{PNNMHVnew}) obeys the collinear limit (\ref{coll-super}).
This observation can be immediately generalised to arbitrary non-MHV amplitudes.
The crucial feature is that all invariants share the same first label $n$, which is
simply replaced by $P$ in the collinear limit.

Finally we remark that the
divergent prefactor in (\ref{coll-super}) originates entirely from the
MHV prefactor $\mathcal{A}^{\rm MHV}_n$, and that $\mathcal{P}_n$ (defined in
(\ref{def-curlyP})) has a finite collinear limit.

\section{Conventional and dual superconformal generators} \label{app-generators}

In this appendix we give the conventional and dual representations of the superconformal algebra. We
begin by listing the commutation relations of the algebra $u(2,2|4)$. The Lorentz generators
$\mathbb{M}_{\a \b}$, $\overline{\mathbb{M}}_{\da \db}$ and the $su(4)$ generators
$\mathbb{R}^{A}{}_{B}$ act canonically on the remaining generators carrying Lorentz or $su(4)$
indices. The dilatation $\mathbb{D}$ and hypercharge $\mathbb{B}$ act via
\be
[\mathbb{D},\mathbb{J}] = {\rm dim}(\mathbb{J}), \qquad [\mathbb{B},\mathbb{J}] = {\rm
hyp}(\mathbb{J}).
\ee
The non-zero dimensions and hypercharges of the various generators are
\begin{align} \notag
& {\rm dim}(\mathbb{P})=1, \qqquad {\rm dim}(\mathbb{Q}) = {\rm dim}(\overline{\mathbb{Q}}) =
\tfrac{1}{2},\qquad {\rm dim}(\mathbb{S}) = {\rm dim}(\overline{\mathbb{S}}) = -\tfrac{1}{2}
\\
&{\rm dim}(\mathbb{K})=-1,\qquad {\rm hyp}(\mathbb{Q}) = {\rm hyp}(\overline{\mathbb{S}}) =
\tfrac{1}{2}, \qquad~ {\rm hyp}(\overline{\mathbb{Q}}) = {\rm hyp}(\mathbb{S}) = - \tfrac{1}{2}.
\end{align}
The remaining non-trivial commutation relations are,
\begin{align} \notag
& \{\mathbb{Q}_{\a A},\overline{\mathbb{Q}}_{\da}^B\}  =  \delta_A^B \mathbb{P}_{\a \da},
   \qquad \{\mathbb{S}_{\a}^A,\overline{\mathbb{S}}_{\da B} \} = \delta_B^A \mathbb{K}_{\a \da},
\\ \notag
& {}[\mathbb{P}_{\a \da},\mathbb{S}^{\b A}] = \delta_{\a}^{\b} \overline{\mathbb{Q}}_{\da}^A,
 \qqquad [\mathbb{K}_{\a \da},\mathbb{Q}^{\b}_{A}] = \delta_{\a}^{\b}
   \overline{\mathbb{S}}_{\da A},
\\ \notag
& {}[\mathbb{P}_{\a \da},\overline{\mathbb{S}}^{\db}_{A}]  =  \delta^{\db}_{\da} \mathbb{Q}_{\a A},
\qqquad [\mathbb{K}_{\a \da}, \overline{\mathbb{Q}}^{\db A}]  =  \delta_{\da}^{\db} \mathbb{S}_{\a}^{A},
\\ \notag
& [\mathbb{K}_{\a \da},\mathbb{P}^{\b \db}] = \delta_\a^\b \delta_\da^\db \mathbb{D} +
\mathbb{M}_{\a}{}^{\b}
 \delta_\da^\db + \overline{\mathbb{M}}_{\da}{}^{\db} \delta_\a^\b,
\\ \notag
& \{\mathbb{Q}^{\a}_{A},\mathbb{S}_\b^B\} =  \mathbb{M}^{\a}{}_{\b}
\delta_A^B + \delta^{\a}_{\b} \mathbb{R}^{B}{}_{A} + \tfrac{1}{2}\delta^{\a}_{\b} \delta_A^B (\mathbb{D}+\mathbb{C}),
\\ \label{comm-rel}
& \{\overline{\mathbb{Q}}^{\da A},\overline{\mathbb{S}}_{\db B}\} = \overline{\mathbb{M}}^{\da}{}_{\db} \delta_B^A  - \delta^{\da}_{\db} \mathbb{R}^{A}{}_{B} + \tfrac{1}{2} \delta^{\da}_{\db}\delta_B^A
(\mathbb{D}-\mathbb{C}).
\end{align}
Note that in writing the algebra relations we are obliged to choose the $su(4)$ chirality of the odd generators. The relations above are valid directly for the dual superconformal generators. For the conventional realisation of the algebra, one should simply swap all $su(4)$ chiralities appearing in the commutation relations.
We now give the generators in both the conventional and dual representations of the superconformal
algebra. We will use the following shorthand notation:
\begin{align}
\partial_{i \alpha \dot{\alpha}} = \frac{\partial}{\partial
x_i^{\alpha \dot{\alpha}}}, \qquad \partial_{i \alpha A} = \frac{\partial}{\partial \theta_i^{\alpha
A}}, \qquad \partial_{i \alpha} = \frac{\partial}{\partial \lambda_i^{\alpha}}\,, \qquad
\partial_{i \dot{\alpha}} = \frac{\partial}{\partial
    \tilde{\lambda}_i^{\dot{\alpha}}}\,, \qquad
\partial_{i A} = \frac{\partial}{\partial \eta_i^A}\,.
\end{align}
We first give the generators of the conventional superconformal symmetry, using lower case
characters to distinguish these generators from the dual superconformal generators which follow
afterwards.
\begin{align}
& p^{\dot{\alpha}\alpha }  =  \sum_i \tilde{\lambda}_i^{\dot{\alpha}}\lambda_i^{\alpha} , & &
k_{\alpha \dot{\alpha}} = \sum_i \partial_{i \alpha} \partial_{i \dot{\alpha}},\notag\\
&\overline{m}_{\dot{\alpha} \dot{\beta}} = \sum_i \tilde{\lambda}_{i (\dot{\alpha}} \partial_{i
\dot{\beta} )}, & & m_{\alpha \beta} = \sum_i \lambda_{i (\alpha} \partial_{i \beta )}
,\notag\\
& d =  \sum_i [\tfrac{1}{2}\lambda_i^{\alpha} \partial_{i \alpha} +\tfrac{1}{2}
\tilde{\lambda}_i^{\dot{\alpha}} \partial_{i
    \dot{\alpha}} +1], & & r^{A}{}_{B} = \sum_i [-\eta_i^A \partial_{i B} + \tfrac{1}{4}\delta^A_B \eta_i^C \partial_{i C}],\notag\\
&q^{\alpha A} =  \sum_i \lambda_i^{\alpha} \eta_i^A, &&   \bar{q}^{\dot\alpha}_A
= \sum_i \tilde\lambda_i^\da \partial_{i A}, \notag\\
& s_{\alpha A} =  \sum_i \partial_{i \alpha} \partial_{i A}, & &
\bar{s}_{\dot\alpha}^A = \sum_i \eta_i^A \partial_{i \dot\alpha}.\notag\\
&c = \sum_i [1 + \tfrac{1}{2} \lambda_i^{\a} \partial_{i \a} - \tfrac{1}{2} \tilde\lambda_i^{\da} \partial_{i \da} - \tfrac{1}{2} \eta_i^A \partial_{iA} ]
\end{align}
We can construct the generators of dual superconformal transformations by starting with the standard
chiral representation and extending the generators so that they commute with the constraints,
\be
(x_i-x_{i+1})_{\a \dot\alpha}  - \lambda_{i\, \a}\, \tilde{\lambda}_{i\, \dot\alpha} = 0, \qquad (\theta_i - \theta_{i+1})_\alpha^A - \lambda_{i \alpha} \eta_i^A = 0.
\ee
By construction they preserve the surface defined by these constraints, which is where the amplitude
has support. The generators are
\begin{align}
P_{\alpha \dot{\alpha}}&= \sum_i \partial_{i \alpha \dot{\alpha}},\\
Q_{\alpha A} &= \sum_i \partial_{i \alpha A}, \\
\overline{Q}_{\dot{\alpha}}^A &= \sum_i [\theta_i^{\alpha A}
  \partial_{i \alpha \dot{\alpha}} + \eta_i^A \partial_{i \dot{\alpha}}],\label{barQfss}\\
M_{\alpha \beta} &= \sum_i[x_{i ( \alpha}{}^{\dot{\alpha}}
  \partial_{i \beta ) \dot{\alpha}} + \theta_{i (\alpha}^A \partial_{i
  \beta) A} + \lambda_{i (\alpha} \partial_{i \beta)}],\\
\overline{M}_{\dot{\alpha} \dot{\beta}} &= \sum_i [x_{i
    (\dot{\alpha}}{}^{\alpha} \partial_{i \dot{\beta} ) \alpha} +
  \tilde{\lambda}_{i(\dot{\alpha}} \partial_{i \dot{\beta})}],\\
R^{A}{}_{B} &= \sum_i [\theta_i^{\alpha A} \partial_{i \alpha B} +
  \eta_i^A \partial_{i B} - \tfrac{1}{4} \delta^A_B \theta_i^{\alpha
    C} \partial_{i \alpha C} - \tfrac{1}{4}\delta^A_B \eta_i^C \partial_{i C}
],\\ \label{DD} D &= \sum_i [-x_i^{\dot{\alpha}\alpha}\partial_{i \alpha \dot{\alpha}} -
  \tfrac{1}{2} \theta_i^{\alpha A} \partial_{i \alpha A} -
  \tfrac{1}{2} \lambda_i^{\alpha} \partial_{i \alpha} -\tfrac{1}{2}
  \tilde{\lambda}_i^{\dot{\alpha}} \partial_{i \dot{\alpha}}],\\
  \label{CC}
C &=  \sum_i [-\tfrac{1}{2}\lambda_i^{\alpha} \partial_{i \alpha} +
  \tfrac{1}{2}\tilde{\lambda}_i^{\dot{\alpha}} \partial_{i \dot{\alpha}} + \tfrac{1}{2}\eta_i^A
  \partial_{i A}], \\
S_{\alpha}^A &= \sum_i [-\theta_{i \alpha}^{B} \theta_i^{\beta A}
  \partial_{i \beta B} + x_{i \alpha}{}^{\dot{\beta}} \theta_i^{\beta
    A} \partial_{\beta \dot{\beta}} + \lambda_{i \alpha}
  \theta_{i}^{\gamma A} \partial_{i \gamma} + x_{i+1\,
    \alpha}{}^{\dot{\beta}} \eta_i^A \partial_{i \dot{\beta}} -
  \theta_{i+1\, \alpha}^B \eta_i^A \partial_{i B}],\\
\overline{S}_{\dot{\alpha} A} &= \sum_i [x_{i \dot{\alpha}}{}^{\beta}
  \partial_{i \beta A} + \tilde{\lambda}_{i \dot{\alpha}}
  \partial_{iA}],\label{fssbarS}\\ \label{KK}
K_{\alpha \dot{\alpha}} &= \sum_i [x_{i \alpha}{}^{\dot{\beta}} x_{i
    \dot{\alpha}}{}^{\beta} \partial_{i \beta \dot{\beta}} + x_{i
    \dot{\alpha}}{}^{\beta} \theta_{i \alpha}^B \partial_{i \beta B} +
  x_{i \dot{\alpha}}{}^{\beta} \lambda_{i \alpha} \partial_{i \beta}
  + x_{i+1 \,\alpha}{}^{\dot{\beta}} \tilde{\lambda}_{i \dot{\alpha}}
  \partial_{i \dot{\beta}} + \tilde{\lambda}_{i \dot{\alpha}} \theta_{i+1\,
    \alpha}^B \partial_{i B}].
\end{align}
We also have the hypercharge $B$,
\begin{equation}
B=\sum_i[- \tfrac{1}{2}\theta_i^{\alpha A} \partial_{i \alpha A} -
  \tfrac{1}{2} \lambda_i^\alpha \partial_{i \alpha} +
   \tfrac{1}{2}\tilde{\lambda}_i^{\dot{\alpha}} \partial_{i \dot{\alpha}}]
\end{equation}
Note that if we restrict the dual generators $\bar{Q},\bar{S}$ to the on-shell superspace they
become identical to the conventional generators $\bar s, \bar q$.

\end{document}